\documentclass[12pt,tightenlines,floats,aps,amsmath,amssymb,nofootinbib,prd,shownopacs,floatfix]{revtex4}

\usepackage{amsmath,amssymb,amsfonts,dcolumn,color,graphicx,graphics,latexsym}
\usepackage[mathscr]{eucal}
\usepackage[latin1]{inputenc}
\usepackage{enumerate}
\usepackage{dsfont}
\newcommand{\be}{\begin{equation}}
\newcommand{\ee}{\end{equation}}
\newcommand{\ba}{\begin{eqnarray}}
\newcommand{\ea}{\end{eqnarray}}

\newcommand{\bmult}{\nopagebreak[3]\begin{multline}}
\newcommand{\emult}{\end{multline}}

\usepackage{ulem}

\newtheorem{Theorem}{Theorem}
\newtheorem{Definition}{Definition}

\newcommand{\rd}{{\rm d}}

\newcommand{\Hil}{\mathcal{H}}
\newcommand{\id}{\mathds{1}}

\newcommand{\sinc}{{\rm sinc}}

\begin{document}

\title{New Loop Quantum Cosmology Modifications from Gauge-covariant Fluxes}

\author{Klaus Liegener\thanks{liegener1@lsu.edu}, Parampreet Singh\thanks{psingh@phys.lsu.edu}}
\affiliation {
 Department of Physics and Astronomy, Louisiana State University,
Baton Rouge, LA 70803}

\begin{abstract}
Loop quantum cosmology is a symmetry reduced quantization of cosmological spacetimes based on loop quantum gravity. While it has been successful in resolution of various cosmological singularities and connecting Planck scale physics to phenomenology, its connection with loop quantum gravity has remained elusive. It is therefore important to integrate more and more features of the full theory into this framework and understand the reliability of physical predictions. In particular, if one wishes to connect the effective Hamiltonian in loop quantum cosmology to an expectation value of the scalar constraint operator in suitable coherent states for the full theory, one has to go beyond the standard setting of loop quantum cosmology. One possibility is to introduce gauge-covariant fluxes, which become necessary because the presence of a finite regularization parameter causes functions build out of the standard discretized variables to be in general not gauge invariant. Following the construction of gauge-covariant fluxes pioneered by Thiemann in \cite{ThiVII00}, we show that the physics of loop quantum cosmology is affected in a non-trivial way. The bounce turns out to be generically asymmetric with a rescaling of the Newton's constant in the pre-bounce branch. Gauge-covariant fluxes result in a higher order quantum difference equation in comparison to loop quantum cosmology. Even the behavior of matter, which behaves innocuously in loop quantum cosmology, is enriched, resulting in an effective non-minimal coupling.  These effects are shown to be common to different choices of regularization parameters.

\end{abstract}

\maketitle

\section{Introduction}
\label{s1}
The search for a well defined and physically viable theory of quantum gravity is as of today an uncompleted task. As a promising candidate for this endeavor, loop quantum gravity (LQG) has matured in the recent decades \cite{Rov04,AL04,Thi07}. This approach is a non-perturbative quantization of General Relativity (GR) in its $3+1$ ADM formulation \cite{ADM62}, which is rewritten in terms of the Ashtekar-Barbero variables in which it takes the form of a gauge theory with a ${\rm SU}(2)$ gauge group \cite{Ash86,Ash87,Bar94}. LQG leads to quantum effects of geometry, such as the discrete spectrum of geometric operators. These are expected to result in fundamental changes in the physical predictions from GR at the Planck scale. An important avenue to understand these effects is the very early universe where pertinent questions are whether discrete quantum geometry effects result in resolution of big bang singularity, and if there is any direct or indirect effect in the physics of the very early universe. Answers to these questions have been explored in loop quantum cosmology (LQC) which is based on techniques of LQG adapted to symmetry reduced cosmological spacetimes \cite{Boj05,AP11}. The high degree of symmetry in cosmological models allows to explore many features of quantum gravity by putting methods of LQG in action without various technical difficulties encountered in full LQG itself. A key result of LQC is 
the existence of a quantum bounce first found in homogeneous and isotropic spacetimes sourced with a massless scalar field \cite{APS06a,APS06b,acs}. In recent years, this result has been extended to various isotropic and anisotropic spacetimes, polarized Gowdy models, and current studies aim to uncover quantum gravity signatures via astronomical observations \cite{as-review}. Lessons from LQC have also proved useful in understanding  the resolution of singularities in symmetry reduced black hole models (see for eg. \cite{gop,aos}).

In the last decade, while LQC has served well as a testbed to extract physics from LQG, the question of whether LQC is the cosmological sector of LQG (or any of its incarnations) has not yet been answered \cite{BF07,Fle15,BEHM16,BEHM17,EV18}. One promising method to get some hints on this issue is to study coherent states in some approximations of LQG which are peaked on semi-classical cosmological space times. Some computations of the expectation values of the scalar constraint motivated from LQG have shown agreement with those in LQC for certain regularizations \cite{AC12,AC14a,ABLS19}. These calculations have been improved, for example by considering regularizations of the scalar constraint which do not assume classical symmetries of Friedmann-Lema$\hat{\i}$tre -Robertson-Walker (FLRW) spacetime \cite{DL17a,DL17b,LL19}. It must however be noted that all these computations have so far been performed in {\it non graph-changing} regularizations of the scalar constraint, in other words on a fixed graph or, more precisely, a cubic lattice. On such a finite lattice, whose spacing in some fiducial coordinate metric is a finite parameter $\epsilon$, certain further requirements must be met in order to ensure that coherent states are peaked over well-behaved classical quantities. We outline these requirements in the following.

Recall that in  LQG one does not quantize the Ashtekar-Barbero variables $(A^I_a(x),E^b_J(y))$  directly in order to avoid operator valued distributions, but instead considers suitable smearings thereof, i.e. the holonomy-flux algebra. While the holonomy of the connection is constructed along an edge $e$, the electric field (triad) gets normally promoted to a flux $E(S)$, i.e. smeared against a two-dimensional face $S$. The electric fields are Poisson commuting, but there arise technical difficulties in determining the Poisson bracket between two fluxes which does not close in an obvious way \cite{ACZ98}. It has been proposed that, to ensure consistency with the quantum algebra, one should instead of the holonomy-flux algebra base the quantization on the Lie algebra of holonomies and vector fields associated with the fluxes. Indeed, by construction the vector fields form a Lie algebra and, being derivatives, satisfy the Jacobi identity \cite{ACZ98}. However, applying this method to compute expectation values via coherent states is not straightforward, as it is not obvious how a coherent state could be peaked over a vector field. There is another resolution of the above problem which is directly useful for coherent state constructions. This method which works at least in the presence of a discrete, fixed lattice  is based on so-called ``covariant'' fluxes -- functions $P(S)$ built out of electric field {\it and} connection for each face $S$ \cite{ThiVII00}. Built in such a way, the non-commutativity between holonomies and electric fields can directly lead to a non-commutativity between the covariant fluxes, hence providing a natural explanation for their non-trivial algebra. 

The approach of gauge-covariant fluxes has the advantage that it is more intuitive and adept to construct coherent states peaked on cosmological spacetimes. Thus providing a promising platform to connect LQC with LQG. Apart from this, a major advantage is that it gives the possibility to address the issue of gauge transformations of the basic variables. Note that while under a gauge transformation the electric field becomes $E^b(x)\mapsto g(x)E^b(x)g(x)^{-1}$, for a finite surface $S$ the standard flux does not transform viably  \cite{ThiVII00}. Functions like a regularized volume, built from these discretized variables are hence a priori not gauge-invariant. Only in the continuum limit $\epsilon\to 0$ of infinitely dense lattices, the gauge-invariance gets restored. But, if one would work in the presence of a finite lattice parameter, one should consider different phase space functions as basic building-blocks to reobtain gauge invariance. For these reasons such modified fluxes have been actively researched in recent years  \cite{FL04,BDOT10,FGZ11,BD13,DG14,CP16,DFG17,FGS18,MMP10}. However, the very first modification was proposed by Thiemann \cite{ThiVII00}, where he constructed the {\it gauge-covariant flux} $P(e)$, such that for the edge $e$ there exists a corresponding face $S$ in the associated dual cell-complex to a well-behaved graph $\Gamma$. Under a gauge transformation the gauge-covariant flux transforms as $P(e)\mapsto g(e(0))P(e)g(e(0)^{-1})$, which allows even in the presence of finite lattice spacing $\epsilon$ to construct gauge invariant observables on the discrete phase space of $\Gamma$.

In contrast to this, in LQC one does not work with fluxes, but with holonomies and the symmetry reduced triad (using classical symmetry properties of FLRW spacetime). Therefore, one avoids the necessity to ask about the gauge transformation properties of any discrete fluxes. Simultaneously, however, one discretizes the connection to holonomies and studies their value along so-called minimal area loops. The resolution of the initial singularity found in LQC is primarily based on the finiteness of this minimal area. In contrast to LQG there is certainly a disparity in the treatment of connection and triads. For one, it would be advantageous to avoid this disparity in the treatment of discretizing connection but not triad during the quantization procedure. Further, one would like to relate the results of LQC somehow to LQG where a discretization of fluxes is intrinsically used. However, finite minimal area loop, i.e. finite discretization means that one should work with gauge-covariant fluxes. For the standard fluxes there exist gauge fixings of the triads to map a discretized set of conventional fluxes  from one classical geometry to any other one (even degenerate ones). It is therefore not possible to physically distinguish these spacetimes \cite{LS19a}. This serious issue must be tackled squarely in order to understand the reliability of LQC predictions vis-\`a-vis the cosmological sector of LQG. Its resolution forces us immediately to work with gauge-covariant fluxes. As we will see, this causes non-trivial modifications to the volume and all related quantities, e.g. the LQC quantum constraint.

Following analysis in \cite{Winkler1,Winkler2,Winkler3,ThiemanComplex,SahThiWin}, the gauge field theory coherent state can be build on a given lattice with a finite spacing $\epsilon$. 
The first application of gauge field coherent states in LQG for $U(1)^3$ on a fixed lattice was performed in \cite{AQG2}. In this manuscript, we will work with ${\rm SU}(2)$ group, and motivated by \cite{ThiVII00} we will consider coherent states peaked on the holonomies $h(e)|_{\rm cos}=h(c)$ and gauge-covariant fluxes $P(e)|_{\rm cos}=P(c,p)$ of isotropic, spatially-flat FLRW spacetime. Here $(c,p)$ denote the symmetry reduced connection and triads which form a canonically conjugate pair in the symmetry reduced gravitational phase space. A similar coherent state, based on a different graph, was constructed in \cite{MMP10}. A computation of the expectation values of the scalar constraint is then expected to yield in first order in the fluctuation the discretized scalar constraint of cosmology, i.e. $C^\epsilon(h(c),P(c,p))$.  This expectation value is  different from standard regularizations in LQC. Employing the conjecture that this discretized constraint can be used as an effective Hamiltonian results in new physics. The task of this manuscript will to be to investigate a loop quantization of cosmology including these {\it gauge-covariant flux corrections} and study a proposal for the effective dynamics of this system. To put this into action for a concrete toy model, we will study a certain regularization of the Hamiltonian constraint which is often used in standard LQC \cite{abl,APS06b,APS06c}. This is based on utilizing the classical symmetry, that in FLRW spacetime the extrinsic curvature is proportional to the connection, using which one combines the Euclidean and Lorentzian term of the Hamiltonian constraint before the quantization process.  Analysis of some properties of the scalar constraint with both the terms discretized independently of each other is performed in a companion paper \cite{LS19c}, and the results of both papers were partially summarized in \cite{LS19a}.

The outline of the paper is as follows. In Sec. \ref{s2}, we discuss how the phase space of a discretization of a continuous manifold can be obtained. This construct follows \cite{ThiVII00} and focuses especially on the fact that all the basic discrete variables transform feasibly under gauge transformation, such that one can easily construct gauge invariant functions from them. This procedure is then applied explicitly in the context of spatially-flat isotropic cosmology, where we adapt the point of view of standard LQC, that is to combine the Euclidean part and Lorentzian part of the scalar constraint using symmetries of FLRW spacetime. Then, a gauge-invariant discretization of the resulting scalar constraint is constructed. This regularization of the scalar constraint will be referred to as standard regularization. Alternatively, one could also treat the Lorentzian part of the scalar constraint independently without invoking any symmetries (also denoted as Thiemann-regularization), however we will fill in on these details in the companion publication \cite{LS19c}. 

The standard regularization (effective dynamics) of the scalar constraint is then studied in Sec. \ref{s3} where the evolution of the isotropic phase space variables $(c,p)$ is investigated. While this has to be done by numerically solving Hamilton's equations, the asymptotic behavior of the system can be computed explicitly in the form of modified Friedmann and Raychaudhuri equations. It transpires that the backward time evolution of a classical universe, resolves the initial singularity, as usual, via a big bounce. However, the evolution in the far past can be matched to a whole family of universes with rescaled constants. All of them differ in a rescaling of momentum of the scalar field and lapse function, but agree in the same, physical observable rescaling of the gravitational constant $G \to \bar G=G(2/\pi)^4$. In this way the big bounce is found to be asymmetric, which is the main feature by which this model deviates from the result in standard LQC. This analysis will be demonstrated for two choices of lattice parameter $\epsilon$: $\mu_0$ (old LQC \cite{abl,APS06b}) and  $\bar{\mu}$-scheme (improved dynamics \cite{APS06c}). Though both choices result in a bounce, their physics has striking differences as in standard LQC. One of the problems of $\mu_0$ scheme in standard LQC is that the bounce occurs at smaller energy densities for larger values of scalar field momentum. Though gauge-covariant fluxes modify the matter Hamiltonian in a non-trivial way, this problem is found to be without resolution. 

In Sec. \ref{s4} we turn towards the quantization of the regularized constraint using methods of LQC. Due to the gauge-covariant flux corrections, the quantization is not straightforward, but can be achieved by expressing the corrections in terms of a infinite series of shift-operators. This leads to an manifestly non-local, yet bounded quantum operator, whose interaction drops sufficiently fast with the LQC-lattice distance.

Finally, we summarize our results in Sec. \ref{s5} and finish with an outlook for further research.

\section{Discrete Symplectic Structures}
\label{s2}
In this section we outline the construction of a discrete phase space for a cubic lattice. The fundamental phase space functions will be built in such a way that they transform covariantly under local ${\rm SU}(2)$-gauge transformations and naturally a non-commuting Poisson brackets between the fluxes arises. All of this is in analogy to the construction of Thiemann in \cite{ThiVII00} to which we refer the reader for all details.

Afterwards we will use his proposal in the explicit context of isotropic, spatially flat cosmology.

\subsection{General Motivation}
We begin by outlining the classical continuum phase space $(M,\Omega)$ of GR in the Ashtekar-Barbero variables on a spatial manifold $\sigma$. Here $M$ denotes the pair of canonical phase variables: connections $A^I_a(x)\tau_I$ and electric fields $E^b_J(y)\tau_J$ respectively, and $\Omega$ is a symplectic structure. $\tau_I = i\sigma_I/2$ ($I=1,2,3$) denote the generators of the Lie algebra $\mathfrak{su}(2)$, satisfying $\mathrm{tr}(\tau_I\tau_J)=-\delta_{IJ}/2$ with
$\sigma_I$ as the Pauli matrices. 
Using 
\be
\Omega=\frac{2}{\kappa\gamma}\int_{\sigma}\rd^3x\; {\rm d}E^a_I(x)\wedge{\rm d}A^I_a(x)
\ee
we can define the 
Poisson bracket through $\{f,g\}:=\Omega(\chi_f,\chi_g)$. Explicitly, for a vector density test field $F^a_I$ of weight one and a covector test field $f^I_a$, one obtains for the smeared quantities
\begin{align}
A[F]:=\int_\sigma \rd^3x\;F^a_IA^I_a\hspace{30pt}{\rm and}\hspace{30pt}E[f]:=\int_\sigma \rd^3x\;E^a_If^I_a ~,
\end{align}
the following Poisson brackets
\begin{align}
\{E[f],E[f']\}=0=\{A[F],A[F']\},\hspace{30pt}\{E[f],A[F]\}=\frac{\kappa\gamma}{2}\int_\sigma \rd^3x\;F^a_If^I_a
\end{align}
where $\kappa=16\pi G$ is the gravitational coupling constant and $\gamma \in\mathbb{R}$ is the Barbero-Immirzi parameter.

We will now consider a truncation of $(M,\Omega)$ to a given graph $\Gamma$. 
The graph $\Gamma$ we consider is a cubic lattice (with possibly infinitely many vertices \cite{SahThiWin}). This graph shall be adapted to a fiducial metric $\eta$ in such a way that the edges of its three directions are along the 3-axes of coordinates. The  
coordinate length of each edge is $\epsilon>0$ with respect to $\eta$. 

Constructed in this way, $\Gamma$ allows the definition of a dual cell-complex of faces which is unique up to diffeomorphisms. We choose its precise form with respect to the fiducial metric $\eta$ in the following way \cite{ThiemanComplex}.  To each edge $e$ of $\Gamma$ we assign an open face $S_e$ carrying the same orientation as $e$ and such that (i) the faces are mutually non-intersecting, (ii) only $e$ intersects $S_e$, (iii) the intersection happens only in one point $e(1/2)$ whose distance to start and end of the edge is both measured as $\epsilon/2$ with respect to $\eta$, and, (iv) if $e$ is oriented along direction $k$ then $S_e$ is oriented along the directions orthogonal to $k$.

Along the lines of \cite{ThiVII00} we define the discrete symplectic manifold $(M_\Gamma,\Omega_\Gamma)$ coming from the continuum $(M,\Omega)$:
\begin{Definition}
i) For any face $S_e$ of the dual cell-complex for $\Gamma$ let $p_0=S_e \cap e$. For any point $x\in S_e$ choose a piecewise analytic path $\rho_x$ such that $\rho_x(0)=p_0$ and $\rho_x(1)=x$. Also, for each edge we call $e_{1/2}\subset e$ the partial path of $e$ with $e_{1/2}(0)=e(0)$ and $e_{1/2}(1)=p_0$.\\
ii) We define the following functions on $(M,\Omega)$: the \emph{holonomy} $h(e)\in {\rm SU}(2)$ of an edge $e$
\begin{align}
h(e)=h(e)(A):=\mathcal{P}\exp\left(\int_0^1\rd t\; A^J_a(e(t))\tau_J\dot{e}^a(t)\right)
\end{align}
and the \emph{gauge-covariant flux} $P^J(e)$:
\begin{align}\label{GaugeCovFlux}
P^J(e)=P^J(e)(A,E):=-2\; {\rm tr}\left(\tau_I h(e_{1/2})\int_S h(\rho_x) \ast E(x) h^{-1}(\rho_x) h^{-1}(e_{1/2}) \right) ~.
\end{align}
\end{Definition}

The first advantage of using functions $h(e), P(e)$ is that their behavior under gauge transformation is well understood. The ${\rm SU}(2)$-Gauss constraint (whose vanishing must be imposed to the restrict the phase space of the Ashtekar-Barbero variables to one equivalent with GR) implies the following Hamiltonian flow on the phase space variables
\begin{align}
A_a(x) \mapsto -(\partial_a g)(x)g^{-1}(x)+g(x)A_a(x)g^{-1}(x)\hspace{20pt}{\rm and}\hspace{20pt}E^a(x) \mapsto g(x)E^a(x)g^{-1}(x)
\end{align}
for any field $g:\sigma\to {\rm SU}(2)$. Using above we find that a holonomy $h(e)$ of the path $e$ transforms as
\begin{align}
h(e)\mapsto g(e(0))h(e) g^{-1}(e(1)) ~,
\end{align}
where $e(0)$ is the beginning point and $e(1)$ is the final point of the path $e$. In contrast to the standard fluxes, the gauge-covariant flux $P(e)$ transforms covariantly
\begin{align}
P^J(e)\tau_J \mapsto g(e(0))P^J(e)\tau_Jg^{-1}(e(1)) ~.
\end{align}
This fact allows the construction of discretized phase space functions, which are gauge-invariant even {\it before} removal of the regulator. As a concrete example, consider a family of lattices $\{\Gamma_\epsilon\}_{\epsilon}$ such that they lie infinitely dense in $\sigma$ for $\epsilon\to 0$. Let $v\in \sigma$ be a vertex of $\Gamma_{\epsilon}$ $\forall \epsilon$ then
\begin{align}
Q^\epsilon(v):=\frac{1}{3!}\sum_{e_1\cap e_2\cap e_3=v}\epsilon(e_1,e_2,e_3)\epsilon_{IJK}P^I(e_1)P^J(e_2)P^K(e_3) \underset{\epsilon\to 0}{\longrightarrow}
\det(E)(v)=Q(v)~. \label{discretisedVolume}
\end{align}
In other words $Q^\epsilon(v)$ is a discretization of the continuum function $Q(v)$ while being gauge-invariant for {\it{all}} $\epsilon$.\footnote{Note that the necessity of implementing a flux that transforms covariantly disappears if one considers everything the limit of infinitesimal faces. Given that the fluxes reduce to the triad itself for vanishing regulator, its gauge-transformation is restored in the continuum limit. Of course, this is exactly what has been done in standard LQG in order to build geometrical operators \cite{RS94,AL96,AL98,Thi98a,Thi98b}.}

The second advantage of using the functions $P(e)$ in contrast to the standard smeared fluxes, is that it was proven in \cite{ThiVII00} how a natural Poisson bracket arises for each $(M_\Gamma,\Omega_\Gamma)$.\footnote{This was done in the presence of regulated tubes around the holonomies and thick surfaces. The brackets are then computed in presence of the regulator, which can afterwards be removed smoothly. We refer the reader to \cite{ThiVII00} for various details.} In particular, we are interested in the following theorem \cite{ThiVII00}:
\begin{Theorem}
	The smeared functions $h(e)(A),P(e)(A,E)$ give rise to the following bracket $\{.,.\}_\Gamma$ on $M_\Gamma$:
	\begin{align}\label{Alg1}
	\{h(e),h(e')\}_{\Gamma}&=0\\
	\{P^I(e),h(e')\}_{\Gamma}&=\frac{\kappa\gamma}{2}\delta(e,e')\tau_I h(e)\\
	\{P^I(e),P^J(e')\}_{\Gamma}&=-\frac{\kappa\gamma}{2}\delta(e,e')\epsilon_{IJK}P^K(e)\label{Alg3}
	\end{align}
	and $\{.,.\}_\Gamma$ satisfies the Jacobi identity and defines a non-degenerate, exact two-form on $M_\Gamma$, that is, it is a symplectic structure.
\end{Theorem}

With the algebra (\ref{Alg1})-(\ref{Alg3}) at hand, canonical quantization on the fixed lattice can now be carried out as is standard in the literature \cite{KS75}. Indeed, following the usual procedure in LQG and promoting holonomies to multiplication operators and gauge-covariant fluxes to right-invariant vector fields ($f\in\mathcal{H}_\Gamma$):
\begin{align}
\hat{h}_{mn}^{(1/2)}(e') f(\{g_e\}_{e\in\Gamma}):=D^{(1/2)}_{mn}(g_{e'})f(\{g_e\}_{e\in\Gamma})\\
\hat{P}^I(e')f(\{g_e\}_{e\in\Gamma}):=\frac{-i\hbar\kappa\gamma}{2}R^K(e')f(\{g_e\}_{e\in\Gamma}),\label{Poperator}
\end{align}
one finds that these satisfy commutation relations which {\it exactly} agree with the canonical quantization rule $i\hbar\{.,.\}\to [.,.]$. Namely,
\begin{gather}\label{quantum_algebra}
[\hat{h}^{(1/2)}(e),\hat{h}^{(1/2)}(e')]=0,\hspace{30pt}[R^K(e),\hat{h}^{(1/2)}(e')]=\delta_{ee'}\tau_K\; \hat{h}^{(1/2)}(e')\\
[R^I(e),R^J(e')]=-\delta_{ee'} \epsilon_{IJK}R^K(e) ~.
\end{gather}
In this framework based on a finite lattice it is hence easy to understand the functions $P^I(e)$ as the semiclassical limit of the operators (\ref{Poperator}), since the commutator algebra reflects the classical Poisson bracket algebra. A coherent state in $\mathcal{H}_\Gamma$ should therefore be chosen such that it is peaked on $P^I(e)$ instead of the standard smeared fluxes $E^I(S)$. This is exactly what has been proposed in \cite{Winkler1,Winkler2,Winkler3,ThiemanComplex,SahThiWin}.

In the following, we will hence adopt this strategy and regularize all physical quantities as functions of $h(e),P^I(e)$, as, e.g., shown in (\ref{discretisedVolume}) for the square of the volume of a single cell. More complicated operators like the scalar constraint can be regularized following a similar strategy \cite{LL19}.

\subsection{Application to Cosmology}
Following the coherent state method of computing expectation values, we will in the this section compute $P^I(e)$ explicitly of a cubic lattice $\Gamma$ with spacing $\epsilon>0$ for isotropic, spatially-flat FLRW cosmology. 

In a certain gauge-fixing the Ashtekar-Barbero variables for an isotropic spatially-flat metric can be expressed as, 
\begin{align}
A^I_a(x)=c\;\delta^I_a,\hspace{40pt}E^a_I(x)=p\;\delta^a_I
\end{align}
where  $|p|= V_o^{2/3} a^2$ and $c=\gamma V_o^{1/3} \dot{a}/\tilde N$ (only for the classical GR). Here $\tilde N$ is the lapse and $V_0:=\int_{\sigma_M}d^3x$ is the coordinate volume of a compact subset $\sigma_M\subset \sigma$, which we will choose to be a torus $T^3$ with $V_0=1$ in the following.  We can think of $(c,p)$ as coordinatizing the subspace of the GR phase space representing spatially-flat, isotropic cosmology with a  reduced symplectic structure: 
\begin{align}
\Omega_{\rm cos}=\frac{6}{\kappa\gamma}\rd p \wedge \rd c
\end{align}
from which the Poisson bracket on the reduced space follows:\footnote{The reader should note, that there exists two conventions in LQG literature for the Poisson-bracket. The one commonly used in LQC reads $\{p,c\}=-\kappa\gamma/6$. The additional minus sign, however, has no consequences for physical quantities, so we will stick to (\ref{cos-PB-bracket}) in the following.}
\be\label{cos-PB-bracket}
\{p,c\}=\frac{\kappa \gamma}{6} ~.
\ee
Computing the standard fluxes for the face in the dual cell-complex to a lattice $\Gamma$ oriented along the coordinate axes with coordinate spacing $\epsilon$ one finds,
\be
h(e_k)=\exp(\int_0^1 \rd t \;c \delta^J_a\tau_J(\epsilon\delta_k^a))=e^{c\epsilon\tau_k}=\cos(\frac{c\epsilon}{2})\id+2\sin(\frac{c\epsilon}{2})\tau_k
\ee
for an holonomy oriented along direction $k$ and
\be\label{StandFluxCos}
E^I(e_k)=\int_{S_{e_k}}(\ast E_I)(x)=\int_{-\epsilon/2}^{\epsilon/2}\rd u\int_{-\epsilon/2}^{\epsilon/2}\rd v \; p\delta^I_k = \epsilon^2 p\delta^I_k.
\ee

In order to compare this with the gauge-covariant fluxes $P(e)$, one has to choose a certain set of paths $\rho$ in their construction (\ref{GaugeCovFlux}). For the moment, we choose for an edge $e_k$ oriented along coordinate direction $k$ to split $\rho_x=\rho_{x,a}\circ \rho'_{x,b}$ where $\epsilon_{kab}=1$. The path $\rho_{x,a}$ starts from the intersection point $\rho_{x,a}[0]=e_k\cap S_{e_k}$ in direction $\pm a$, stays in $S_{e_k}$ and its tangent vector remains constant. The end point of this path, i.e. $\rho_{x,a}[1]$, agrees with the starting point of $\rho'_{x,b}$ and is chosen such that the latter path has a constant tangent vector oriented along $\pm b$ and ends in point $x$, i.e. $\rho'_{x,b}[1]=x$.

Let $k > 0$ and $\epsilon_{kab}=1$. Then 
\begin{align}
P^I(e_k)=&-2 \, \mathrm{tr} \bigg(\tau_I[\cos(\frac{c\epsilon}{4})\id +2\sin(\frac{c\epsilon}{4})\tau_{k}]\int_{-\epsilon/2}^{\epsilon/2}\rd u_a\int_{-\epsilon/2}^{\epsilon/2}\rd u_b\;h(\rho_{x(u_a,u_b),a})\times\\
&\hspace{5pt}\times h(\rho'_{x(u_a,u_b),b})\epsilon_{abd}E^d(x(u_a,u_b))h(\rho'_{x(u_a,u_b),b})^{-1} h(\rho'_{x(u_a,u_b),a})^{-1}[\cos(\frac{c\epsilon}{4})\id +2\sin(\frac{c\epsilon}{4})\tau_k]\bigg)\nonumber
\end{align}
where we choose a coordinate system of $S_{e_k}$ to parametrize the points $x=x(u_a,u_b)$. Note that
\ba\label{innermostpath}
\int_{-\epsilon/2}^{\epsilon/2}\rd u_b\;h(\rho'_{x(u_a,u_b),b})\epsilon_{abd}E^d(x(u_a,u_b))h(\rho'_{x(u_a,u_b),b})^{-1}&&\nonumber\\
 && \hspace{-8cm}=  \int_{-\epsilon/2}^{\epsilon/2}\rd u_b[\cos(\frac{cu_b}{2})\mathds{1}+2\sin(\frac{cu_b}{2})\tau_J]p\tau_K\delta^K_k[\cos(\frac{cu_b}{2})\mathds{1}-2\sin(\frac{cu_b}{2})\tau_J]\nonumber \\
&& \hspace{-8cm}= \int_{-\epsilon/2}^{\epsilon/2}\rd u_b\; p\;\delta^K_k[\cos^2(\frac{cu_b}{2})\tau_K -4\sin^2(\frac{cu_b}{2})\tau_J\tau_K\tau_J]=\int_{-\epsilon/2}^{\epsilon/2}\rd u_b\;p\;\delta^K_k[\cos(cu_b)]\tau_K \nonumber\\
&& \hspace{-8cm}= p\;\delta^K_k\tau_K\frac{\sin(c\epsilon/2)}{c/2}
\ea
where we used the vanishing of odd functions under the integral,  $\tau_J\tau_K\tau_J=\tau_K/4$ if $J\neq K$, and $\cos^2(x)-\sin^2(x)=\cos(2x)$. As the result of (\ref{innermostpath}) is again proportional to $\tau_k$, the same calculation goes through for the integral over $u_a$. Finally, using $\tau_I^2=-1/4$, $-2 \mathrm{tr}(\tau_I\tau_J)=\delta_{IJ}$ and $\mathrm{tr}(\tau_I)=0$ we obtain
\begin{align}
P^I(e_k)&=-2 \, \mathrm{tr}\left(\tau_I[\cos(\frac{c\epsilon}{4})\id+2\sin(\frac{c\epsilon}{4})\tau_k]\tau_k[\cos(\frac{c\epsilon}{4})\id-2\sin(\frac{c\epsilon}{4})\tau_k]\right)p\frac{\sin^2(c\epsilon/2)}{(c/2)^2}\nonumber\\
&=[\delta^I_k\cos(\frac{c\epsilon}{4})^2+\delta^I_k\frac{1}{4}4\sin^2(\frac{c\epsilon}{4})]p\frac{\sin^2(c\epsilon/2)}{(c/2)^2}=(\epsilon^2 p\;\delta^I_k)\; \frac{\sin^2(c\epsilon/2)}{(c\epsilon/2)^2}\nonumber\\ 
&=E^I(e_k)\; \sinc^2(c\epsilon/2) ~.
\end{align}
The remarkable property of this result is, that it allows us to pass from the known expressions in the literature using the standard fluxes directly to the gauge-covariant fluxes by simply replacing $p\to p\;\sinc^2(c\epsilon/2)$. One can hence avoid repetition of lengthy computation of expectation-values of coherent states, and instead immediately obtain results by using the gauge-covariant flux in expressions of \cite{DL17a,DL17b,LL19}. In other words, the leading contribution will be nothing else but the regularization of the scalar constraint in the presence of the gauge-covariant fluxes.\\

Motivated by these findings, we will close this section by giving an explicit example, of how such a gauge-invariant regularization looks like.  For the purpose of this example, we choose the scalar constraint where the Lorentzian part is put proportional to the so-called Euclidean part.\footnote{One might take the point of view, that we restrict partially to cosmology before discretization. Such a restriction leads to this symmetry, as the curvature term of the scalar constraint vanishes.} Moreover, we add a homogeneous, massless, free scalar field $\phi$ minimally coupled to gravity, which can play the role of a relational clock. The Hamiltonian constraint in this case is given by,
\begin{align}
C[\tilde N]=\int \rd^3x\; \tilde N\left(\frac{1}{\kappa\gamma^2}\frac{\epsilon^{IJK} E^a_JE^b_K}{\sqrt{|\det(E)|}}F^I_{ab}+\frac{\pi_\phi^2}{2\sqrt{|\det(E)|}}\right)
\end{align}
where $\pi_\phi$ is the canonical conjugated momentum to $\phi$.

We use the standard regularization strategy from \cite{Thi98a,Thi98b}, where the curvature of the connection gets approximated by a loop $\Box^\epsilon$ and the volume of a small region by $\sqrt{|Q^\epsilon|}$ from (\ref{discretisedVolume}). Using the homogeneity symmetry of the cosmological model, this leads finally to the regularized expression:
\ba
C^\epsilon[\tilde N] &=& \frac{-4 \tilde N}{48\kappa^2\gamma^3}\sum_{ijk}\epsilon(i,j,k)\mathrm{tr}\left((h(\Box^\epsilon_{ij})-h(\Box^\epsilon_{ji})h(e_k)\{h(e_k^{-1},\sqrt{|Q^\epsilon|}\}\right)+\frac{\tilde N \pi_\phi^2}{2\sqrt{|Q^\epsilon|}}\nonumber \\
&=&-\frac{6\tilde N}{\kappa\gamma^2}\frac{\sin^2(c\epsilon)}{\epsilon^2}\sqrt{|p|}\;\sinc(c\epsilon/2)+\frac{\tilde N\pi_\phi^2}{2\sqrt{|p^3|}}\sinc^{-3}(c\epsilon/2)\label{Reg_Constraint_E}~.
\ea
Like any suitable regularization it holds that $\lim_{\epsilon\to 0}C^\epsilon[\tilde N]=C[\tilde N]$ and the deviation from earlier regularizations considered in the literature \cite{APS06a,APS06b,APS06c} is in the aforementioned presence of the $\sinc$-terms. It is the impact of these non-trivial corrections which we will study in the further sections of this article.

We emphasize that this {\it classical} discretization can serve as the starting point of a loop quantization, where the holonomy-flux algebra gets promoted to operators along (\ref{quantum_algebra}). In order to ensure normalizability, the associated Hilbert space is then chosen as the set of possibly all finite graphs. For example the subset of cubic graphs with periodic boundary condition can represent a suitable discretization of the compact torus $\sigma_M\subset \sigma$. 

\section{Regularized Dynamics for gauge-covariant fluxes}
\label{s3}
The Hamiltonian constraint (\ref{Reg_Constraint_E}) incorporates modifications due to the gauge-covariant fluxes for the case of a massless scalar field coupled to the gravity in a spatially flat FLRW spacetime. For the purpose of this analysis we will fix the orientation of the triad to be positive throughout this section. Before we can understand the resulting modified dynamics, we need to fix the regulator $\epsilon$ of the lattice which is finite in our analysis.
We will investigate two choices which have been extensively studied in standard LQC: $\mu_0$ \cite{abl,APS06b} and $\bar \mu$ schemes \cite{APS06c}. For the first choice, the regulator is a constant which is fixed by comparing the smallest kinematical area enclosed by the holonomies to the minimum non-zero eigenvalue of the area operator in LQG, while for the second choice one uses the physical area of the loop which makes $\bar \mu$ a function of triad $p$ (in particular $\bar \mu$ is proportional to $1/\sqrt{p}$).  In standard LQC, it turns out that both choices result in replacing the big bang singularity by a quantum bounce resulting in a symmetric evolution in pre- and post-bounce regimes. Despite this similarity, there are striking differences in physics for the two choices of regulators. In particular, when using the $\mu_0$ regulator the bounce can be at arbitrarily small (or large) values of the energy density by choosing a sufficiently large (or small) value of $\pi_\phi$. This is in contrast to $\bar \mu$ regulator where the quantum bounce always occurs at a universal value $\rho_{\mathrm{b}} \approx 0.41 \rho_{\mathrm{Pl}}$. Unlike the $\bar \mu$ scheme, the density at the bounce is not independent of the fiducial volume $V_0$ in the $\mu_0$ scheme. Thus, by changing $V_0$ one can induce ``quantum gravitational effects'' even at classical scales. Finally, in presence of a positive cosmological constant the $\mu_0$ scheme leads to a recollapse of the scale factor at large volumes which is in direct contradiction with the cosmological dynamics. It is to be noted that 
these limitations of $\mu_0$ scheme are shared by various other potential choices of regulators which depend on phase space functions, and it is only the $\bar \mu$ scheme which yields physically viable dynamics in standard LQC \cite{cs08}.

Despite these issues with the $\mu_0$ scheme in standard LQC, it is worth investigation for the following reasons. Since the implementation of gauge covariant fluxes changes geometry-  and matter-part of the scalar constraint non-trivially it is not clear if such a recollapse  would still occur. Indeed, the properties of a system with gauge covariant fluxes will be vastly different than before and uncover many unexpected features as we will discuss below. Moreover, when restricting the $\mu_0$ scheme to a compact torus, the dependence of the coordinate volume accounts merely to a diffeomorphism dependency \cite{EV18}. Therefore to understand restrictions on any regularization one needs to carefully study diffeomorphism invariant observables. Lastly, it must be emphasized that the $\mu_0$ scheme of cosmology is the only one which can be obtained as a reduction of a gauge-invariant regularization of the general connection and triad fields. Until the present day, the $\bar{\mu}$-scheme lacks such a relation to the full theory.

In contrast to standard LQC, obtaining reliable dynamics in the presence of gauge-covariant fluxes is far more involved due to the complexity of the dynamical equations. Though obtaining a closed form for modified Friedmann and Raychaudhuri equations is quite difficult, in order to understand resulting dynamics we obtain the modified Hamilton's equations  resulting from (\ref{Reg_Constraint_E}) which are then numerically solved.\footnote{As is usual in LQC for spatially flat FLRW model, we ignore modifications resulting from quantization of inverse triad operators to regularized dynamics. In standard LQC, such modifications result in negligible effects in dynamical evolution \cite{APS06b,APS06c}.}  In LQC literature, such dynamics is considered to be the effective dynamics which in various cases has been rigorously confirmed using numerical simulations of the coherent states evolved using quantum evolution operator (see for eg. \cite{APS06c,numlsu-1,numlsu-2}). The effective dynamics in LQC turns out to be in perfect agreement with the expectation value of the quantum Hamiltonian constraint of the same coherent states \cite{Taveras}. Assuming that a similar effect may hold also true for models involving gauge-covariant flux corrections, we will later in Sec. \ref{s4}  construct a quantum operator, whose expectation value is expected to be (\ref{Reg_Constraint_E}) in leading order in the spread of the state. In our analysis, we refer to the evolution produced by it as ``regularized dynamics". From this regularized dynamics, the main result is that the big bang singularity is replaced by a quantum bounce but there are many significant changes from standard LQC as will be described below. We further note that by  incorporating the gauge-covariant fluxes, we have introduced a regularization of the triad $p \to p_{\rm g.c.}=p \; {\rm sinc}^2(\epsilon c/2)$. Similarly, the volume becomes $v_{\rm g.c.} := p^{3/2}{\rm sinc}^3(\epsilon c/2)$. As a result the energy density using regularized volume is given by 
 $\rho = {\cal H}_M/v_{\rm g.c.}$ respectively, where  ${\cal H}_M$ is the matter Hamiltonian. 

Above non-trivial change in the energy density seemingly implies that the matter energy conservation, which is true in standard LQC, may no longer hold. However, it can be shown that the energy conservation does hold with gauge-covariant flux modifications, if one carefully takes into account, the time evolution of energy density and the expansion rate.

\subsection{The $\mu_0$-scheme}
This choice of the regulator appeared in the earliest works in LQC, and is based on considering the minimum square area of the loops over which holonomies are constructed, with respect to the fiducial metric, and equating it with the minimum non-zero eigenvalue $\Delta=4\sqrt{3}\pi\gamma\ell^2_P$  of the area operator in LQG \cite{abl,Boj05,APS06a,APS06b} which results in: 
\be\label{mu0scheme}
\epsilon=\mu_0,\hspace{30pt}\mu_0:=
3\sqrt{3} ~
\ee
In the following, we will choose natural units $\ell_P=G=\hbar=c=1$. 

The vanishing of the Hamiltonian constraint (\ref{Reg_Constraint_E}) immediately results in the following expression of energy density:
\be
\rho=\frac{6}{\kappa \gamma^2\mu_0^2} p^{1/2} \sin^2(c \mu_0)\sinc^{-2}(c \mu_0/2) ~.
\ee
From this expression we see that there is no global bound for the energy density in the dynamical evolution in the $\mu_0$ scheme (as is the situation in standard LQC). Depending on the value of the triad $p$ at which the quantum bounce occurs, which in turn is determined by the value of $\pi_\phi$, the energy density at the bounce changes and can be much smaller or larger than the Planckian value.

\begin{figure}[tbh!]
	\begin{center}
		\includegraphics[scale=0.45]{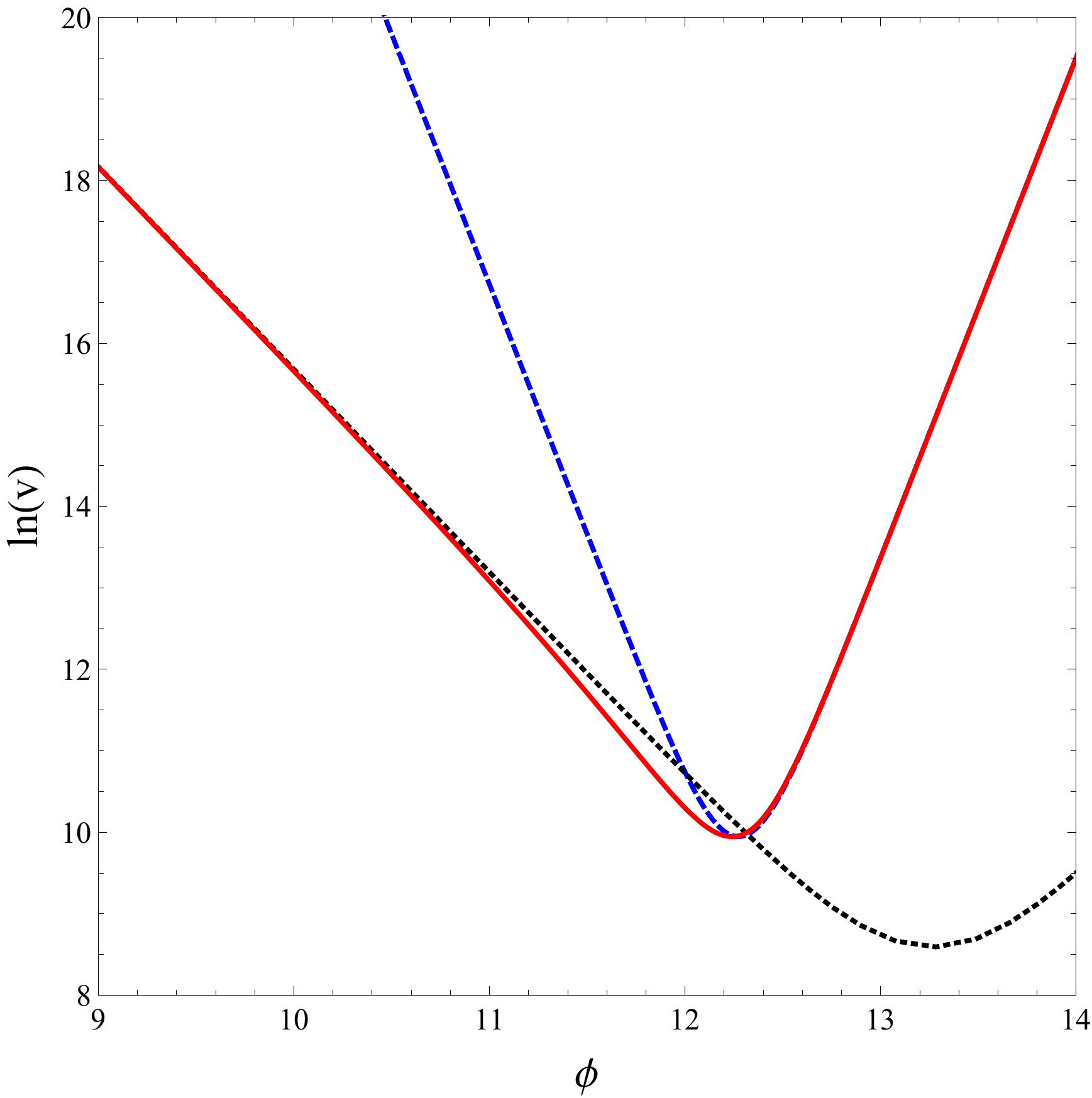}
		\includegraphics[scale=0.45]{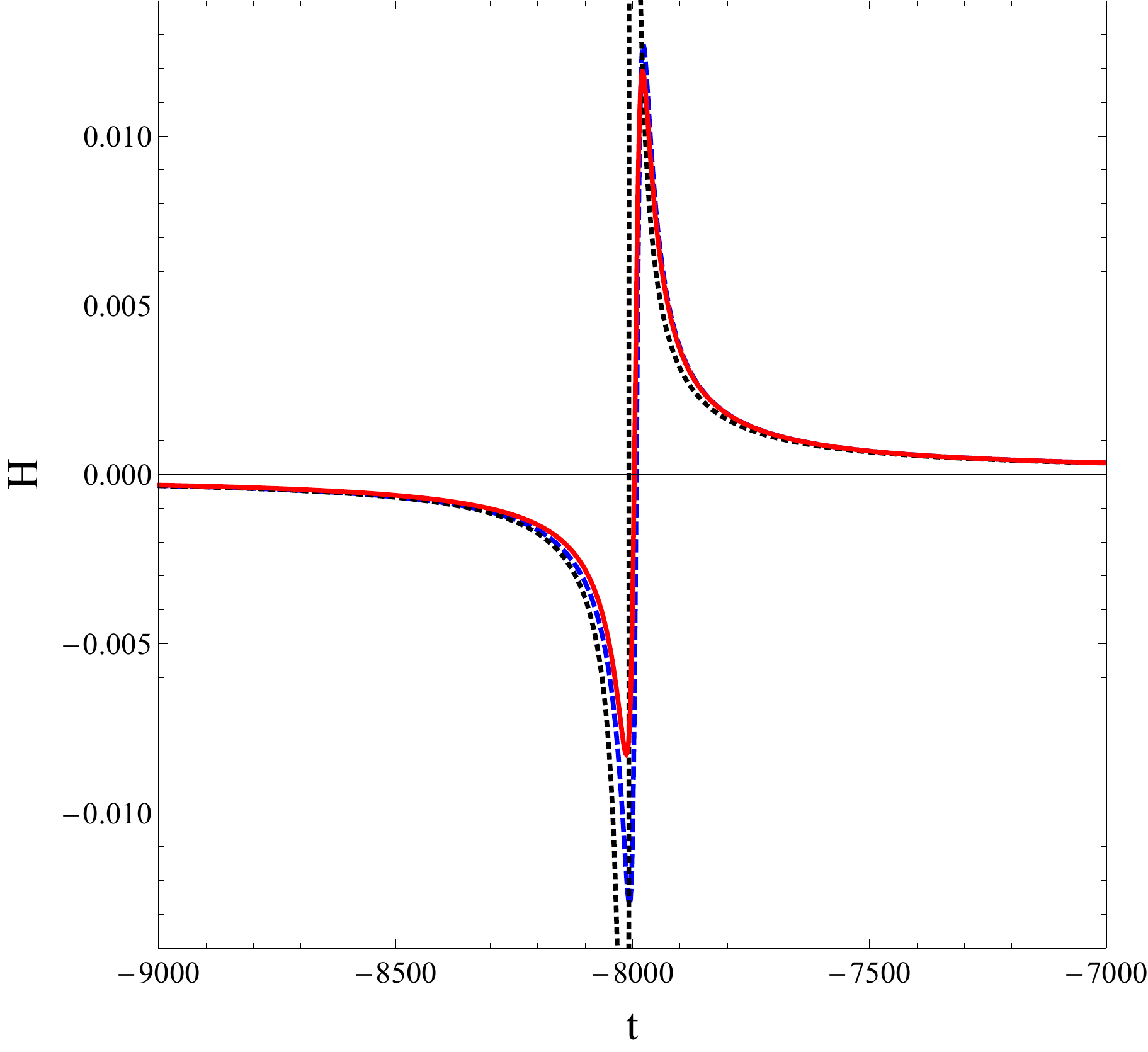}
	\end{center}
\caption{Plots show the evolution of gauge-covariant volume with respect to scalar field, and Hubble rate (computed using gauge-covariant scale factor) with respect to coordinate time $t$ for the case of Hamiltonian constraint (\ref{Reg_Constraint_E}) (red-solid curve). Comparison is made to LQC solution (blue-long dashed curve)  starting from same initial conditions in the far future using standard volume $v$. Lapse is chosen to be unity for these simulations. The black-dashed curve corresponds to LQC solution with a modified values of Newton's constant and scalar field momentum given by (\ref{rescaling}). Note that this the is independent of the choice of $\alpha$. In contrast to standard LQC, gauge-covariant fluxes result in an asymmetric bounce of the universe. In the post-bounce regime there is an excellent agreement with LQC at late times, whereas in the pre-bounce regime the LQC solution with rescaled $G$ and $\pi_\phi$ matches the one with gauge-covariant fluxes modifications. \label{mu0plots}  }  
\end{figure}

The Hamilton's equation can be computed using (\ref{Reg_Constraint_E}) which turn out to be (when choosing lapse $\tilde N=1$)
\ba
\dot{c}&=&-\frac{1}{2\sqrt{p}\gamma}\frac{\sin^2(c\mu_0)}{\mu_0^2}\sinc(c\mu_0/2)-\frac{\pi_\phi^2 \gamma\kappa}{8p^{5/2}}\sinc^{-3}(c\mu_0/2) ~,\\
\dot{p}&=&\frac{1}{16 \mu_0 \gamma  p^{3/2} \sinc^2\left(\frac{c \mu _0}{2}\right)} \Bigg[8 p^2 \cos \left(\frac{c \mu _0}{2}\right) \text{sinc}^4\left(\frac{c \mu _0}{2}\right){} \left(c \mu _0-2 \sin \left(c \mu _0\right)+5 c \mu _0 \cos \left(c \mu _0\right)\right) \nonumber \\ && ~~~~~ +\gamma ^2 \kappa  \mu _0^2 \pi_\phi^2 \left(c \mu _0 \cot \left(\frac{c \mu _0}{2}\right)-2\right) \csc \left(\frac{c \mu _0}{2}\right) \Bigg] ~,\\
\dot{\phi}&=&\frac{\pi_\phi}{p^{3/2}}\sinc^{-3}(c\mu_0/2) ~,\\
\dot{\pi}_\phi&=&0 ~.
\ea
These equations can be numerically solved, a representative of which are plotted in Fig. \ref{mu0plots}. From these plots we see a bounce of the volume of the universe along with boundedness of $\mu_0 c$, $\rho$ and Hubble rate $H$. 
Before we discuss these plots, and differences from standard LQC, it is instructive to study the  asymptotic behavior of $C^{\mu_0}[\tilde N]$ in post-bounce and pre-bounce phases. In particular, we wish to understand the form of the (modified) Friedmann equation and Raychaudhuri equation far away from the bounce, at very large volumes or equivalently very small energy density. 

As in standard LQC, it is clear from $C^{\mu_0}[ \tilde N]=0$ that $\rho$ will become small around the zeros of $\sin(c\mu_0)$. Choosing some of those, we expand around $c=0,\pi/\mu_0$.\footnote{As the numerical investigation below demonstrates, it suffices to study the expansion around these two points. Given suitable initial conditions, the dynamics drives exactly to these asymptotes.} If one chooses one of these points to correspond to post-bounce regime, the other corresponds to the pre-bounce regime (see Fig. \ref{mu0plots}). Around $c\simeq 0$ the scalar constraint from (\ref{Reg_Constraint_E}) becomes (neglecting all contributions higher than $c^3$)
\begin{align}
	C^{\mu_0}[\tilde N]\overset{c\simeq 0}{=}\frac{6\tilde N}{\kappa\gamma^2}\sqrt{p}c^2+\frac{\tilde N\pi_\phi^2}{2p^{3/2}}(1+\frac{c^2\mu_0^2}{8})+\mathcal{O}(c^4)\approx 0 ~.
\end{align}
This can be solved for $c$ as:
\be\label{mu0Asymptotic_c}
	c=\pm \frac{\pi_\phi\sqrt{\kappa\gamma^2}}{p\sqrt{12}}\sqrt{1-\frac{\mu_0^2\pi_\phi^2\kappa\gamma^2}{96p^2}}^{-1} ~.
\ee 
We can then obtain the following Hamilton's equation in the limit of large $p$:
\begin{align}
\dot{p}=\{C^{\mu_0}[\tilde N],p\}=\tilde N\frac{\sqrt{p}}{\gamma}\frac{\sin(2c\mu_0)}{\epsilon}\sinc(c\mu_0/2)+\mathcal{O}(p^{-3/2}) ~.
\end{align}
We eliminate therein $c$ using (\ref{mu0Asymptotic_c}) to obtain the Friedmann equation. Similarly we can use $\ddot{a}=\{C[\tilde N],\dot{p}/(2\sqrt{p})\}$ to determine the Raychaudhuri equation. Together they read:
\begin{align}\label{Friedmann_classical}
\left(\frac{\dot{a}}{a}\right)^2|_{c\simeq 0}&=\tilde{ N}^2\frac{\kappa}{6}\frac{\pi_\phi^2}{2p^3}+\mathcal{O}(p^{-4}) ~,\\
\left(\frac{\ddot{a}}{a}\right)|_{c\simeq 0}&=-\tilde N^2\frac{\kappa}{3}\frac{\pi_\phi^2}{2p^3}+\mathcal{O}(p^{-4}) ~.
\end{align}
As it turns out this agrees in its leading order in the matter energy density with the classical Friedmann and Raychaudhuri equations, which are commonly written with $\tilde N=1$. Note that in the above equations, since $\sinc(c \mu_0)|_{c\simeq 0} \simeq 1$ one gets $p_{\rm{g.c.}} \simeq p$. As a  result, the energy density $\rho = {\cal H_M}/p^{3/2}_{\rm{g.c.}}$ for the massless scalar field equals the conventional expression $\pi_\phi^2/2 p^3$. Hence, we find that in the presence of gauge-covariant flux modifications the asymptotic regime in the neighborhood of $c \simeq 0$ agrees with the one of classical GR.

As already mentioned, another obvious phase space point for small $\rho$ is at $c=\pi/\mu_0$. To determine the effective Friedmann equation at this point, we perform a canonical transformation $c^*:=\pi/\mu_0-c$ and proceed again as before:
\begin{align}
	C^{\mu_0}[\tilde N]\overset{c^*\simeq 0}{=}-\frac{6\tilde N}{\kappa\gamma^2}{c^*}^2\sqrt{p}\frac{1-(c^*\mu_0/2)^2}{\pi/2-c^*\mu_0/2}+\frac{\tilde N\pi_\phi^2}{2p^{3/2}}\left(\frac{1-(c^*\mu_0/2)^2/2}{\pi/2-c^*\mu_0/2}\right)^{-3}+\mathcal{O}({c^*}^3)\approx 0
\end{align}
which can be solved for $c^*$ as:
\begin{align}
c^*=\pm \frac{\pi_\phi\sqrt{\kappa\gamma^2}}{p\sqrt{12}}\left(\frac{\pi}{2}\right)^2+\mathcal{O}(p^{-2}) ~.
\end{align}
We see, that this is already at leading order different than (\ref{mu0Asymptotic_c}), which is caused by the expansion of $\sinc$ around $\pi/2$.

Once again, this will be inserted into $\dot{p}$ as determined by the Hamiltonian flow of the constraint. In the same manner, we determine $\ddot{a}$ via the flow of the constraint and after some algebra we find {\it modified} Friedmann and Raychaudhuri equations:
\ba\label{Friedmann_modified}
\left(\frac{\dot{a}}{a}\right)^2|_{c\simeq \pi/\mu_0}&=&\tilde N^2\frac{\kappa}{6}\frac{\pi_\phi^2}{2p^3}\sinc^{-2}(\pi/2)+\mathcal{O}(p^{-4})=\tilde N^2\frac{\kappa}{6}\frac{\pi_\phi^2}{2p^3}\left[\frac{\pi}{2}\right]^2+\mathcal{O}(p^{-4}),\\
\left(\frac{\ddot{a}}{a}\right)|_{c\simeq \pi/\mu_0}&=&-\tilde N^2\frac{\kappa}{3}\frac{\pi_\phi^2}{2p^3}\sinc^{-2}(\pi/2)+\mathcal{O}(p^{-4})=-\tilde N^2\frac{\kappa}{3}\frac{\pi_\phi^2}{2p^3}\left[\frac{\pi}{2}\right]^2+\mathcal{O}(p^{-4}) ~.
\ea
These equations are manifestly different from the classical expression. Moreover, one realizes that unlike the case of $c=0$,  the velocity of the scalar field is modified around $c=\pi/\mu_0$, namely
\begin{align}
	\dot{\phi}|_{c\simeq\pi/\mu_0}=\{C^{\mu_0}[\tilde N],\phi\}|_{c\simeq\pi/\mu_0}=\tilde N\frac{\pi_\phi}{\sqrt{p}^3}\left[\frac{\pi}{2}\right]^3+\mathcal{O}(p^{-4}) ~.
\end{align}
Above equations imply that one can find a rescaling of $\tilde N$, $\pi_\phi$ and $G$ as
\begin{align}\label{rescaling}
\bar{\pi}_\phi=\pi_\phi\alpha,\hspace{50pt}\bar{N}=\tilde N \left[\pi/2\right]^3 \alpha^{-1},\hspace{50pt}\bar{G}=G \left[2/\pi\right]^4
\end{align}
such that for any $\alpha\in\mathbb{R}$\textbackslash$\{0\}$ the dynamical equations take again the form of classical Friedmann equations with rescaled scalar field momentum  $\bar{\pi}_\phi$ and with modified Newton's coupling constant $\bar{G}$ and new lapse $\bar{N}$. The freedom in choice of $\alpha$ implies that one could for example absorb the rescaling of lapse completely in $\pi_\phi$ . On the other hand, it also possible to choose $\alpha=1$, i.e. we merely rescale the lapse function, which reflects simply a choice of coordinate system and has therefore no physical relevance.\footnote{This can be also seen, when realizing that relational observables, such as $v(\phi)$, do classically not depend on $\pi_\phi$, which is seen therefore to be pure gauge in the $v-\phi$ plane.} 
Thus, if one chooses $\alpha=1$, one isolates the physically relevant rescaling of $G\to\bar{G}$.

\begin{figure}[tbh!]
	\begin{center}
		\includegraphics[scale=0.45]{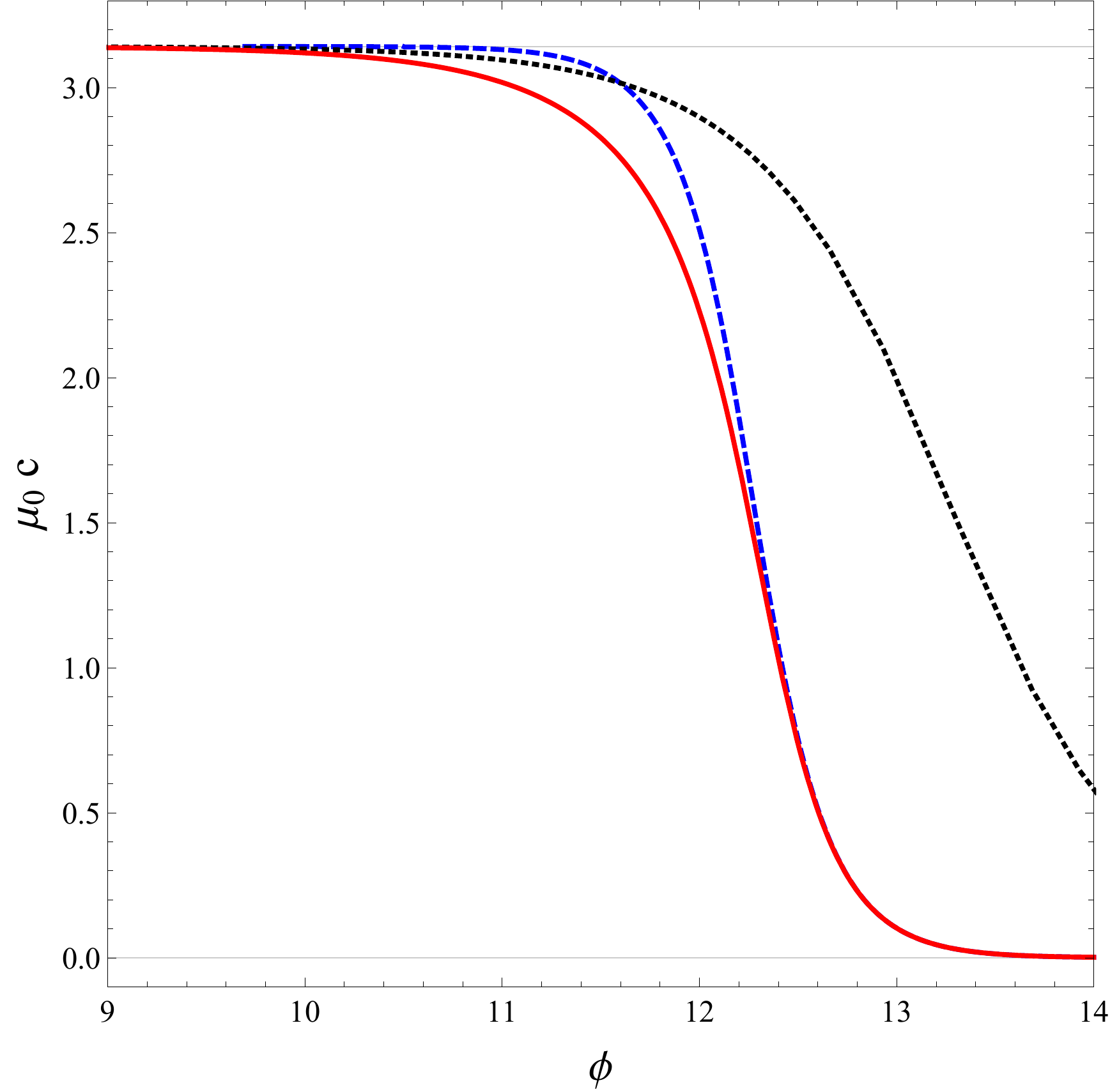}
		\includegraphics[scale=0.45]{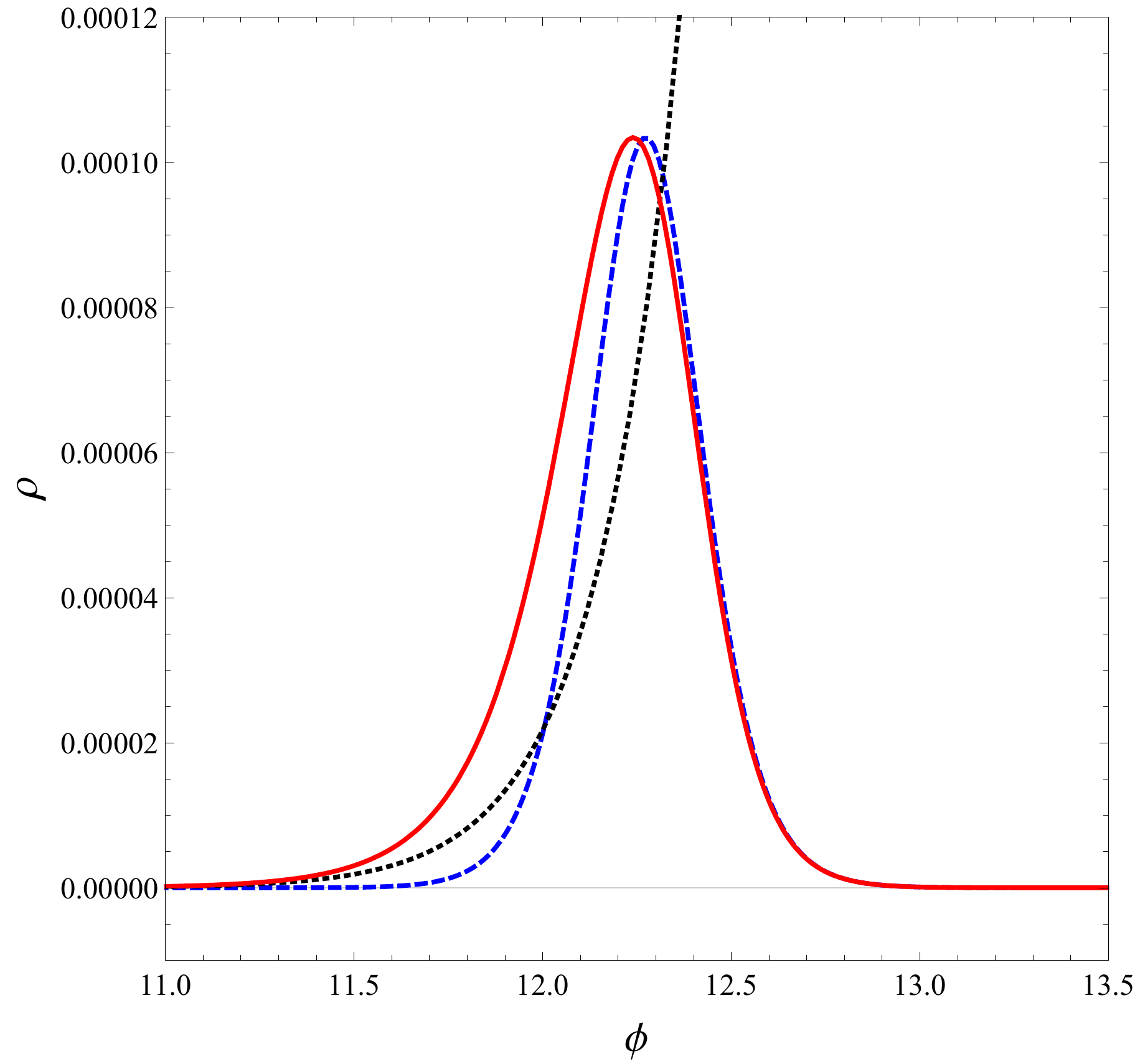}
	\end{center}
\caption{Behavior of connection $c$ and energy density $\rho$ are plotted versus $\phi$. Initial conditions and conventions correspond to those in Fig. \ref{mu0plots}. The black-dashed curve in energy density plot is bounded above by $0.0016$ for the values specified in the main text. This value differs from standard LQC value because of rescaling of the $G$. \label{fig2}}   
\end{figure}

Given the observational constraints on $G$, this asymptotic regime  is ruled out to represent the post-bounce expanding branch of our universe, and can only correspond to the pre-bounce phase (see Ref. \cite{LSW18a} for a discussion on a similar constraint). Thus, in the asymptotic pre-bounce regime, the dynamical equations from   (\ref{Reg_Constraint_E}) correspond to classical Friedmann and Raychaudhuri equations expressed in standard scale factor (or triads) with a modified Newton's coupling constant given by (\ref{rescaling}).   

 We want to point out, that there also exists a second branch of solutions, which are the time-reversed version of the former. If one expands the Friedmann equations around the point $c=-\pi/\mu_0$, one finds another classical asymptote with the same rescaling. However, this solution has to be interpreted in such a way that the rescaled universe lies in the post-bounce branch. Of course, the previous analysis can in principle be repeated for all $b:=c\bar{\mu}\simeq z\pi$ with $z\in\mathbb{Z}$. In standard LQC these choices corresponded to several branches, which were all physically indistinguishable as the flow of the Hamiltonian constraint in standard LQC interpolates between $(z-1)\pi$ and $z\pi$. In that case each point $z\pi$ corresponds to an asymptotic point where classical FLRW universe gets approached. As a result, in standard LQC it is not possible to decide the corresponding branch of the universal dynamics from physical observations.

However, in presence of the gauge-covariant fluxes the situation changes dramatically. Similar to (\ref{Friedmann_modified}), at $c\mu_0\simeq z\pi$ the rescaling in the Friedmann and Raychaudhuri equations goes as $\sinc(z\pi/2)^{-2}$. In other words, the {\it only} two branches featuring as classical FLRW universe in one asymptotic limit are the two principal branches $c\in (-\pi/\mu_0,0)$ and $c\in (0,\pi/\mu_0)$. It should be noted that one principal branch is exactly the time reverse version of the other principal branch. That means that if initial conditions are given in the pre-bounce branch, then a rescaling of constants occurs for the post-bounce branch. The asymptotic properties of the solutions, as discussed above, show that the rescaling  $G\to \bar G \neq G$ is an intrinsic feature of the Hamiltonian with gauge-covariant flux modifications. This rescaling holds irrespective of the initial conditions. Thus, there is  no choice for $G$ such that the effective Newton's constant as measured on both sides of the bounce is the same. The same result holds true for the $\bar \mu$-scheme. 


\begin{figure}[tbh!]
	\begin{center}
		\includegraphics[scale=0.45]{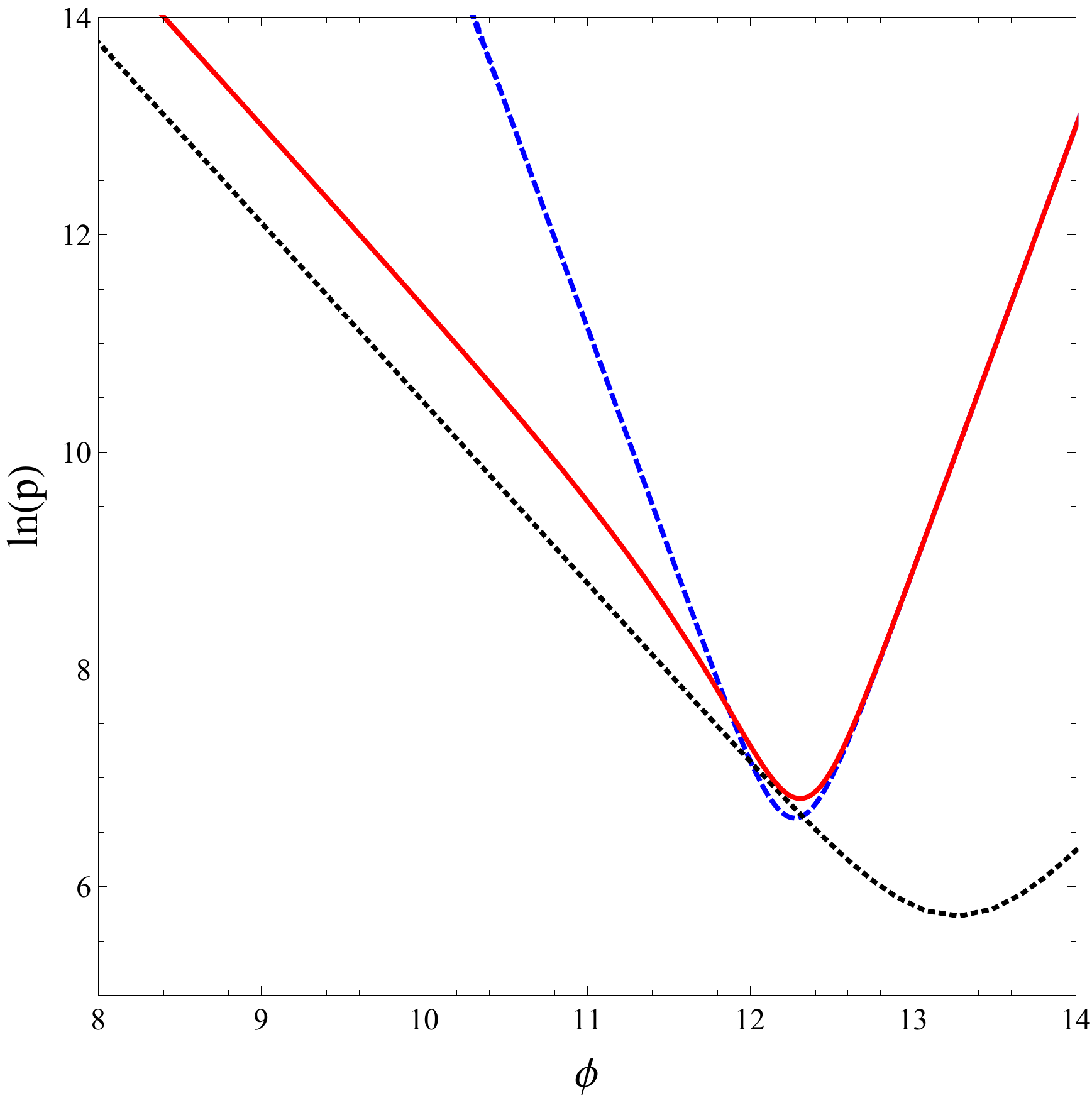}
		\includegraphics[scale=0.53]{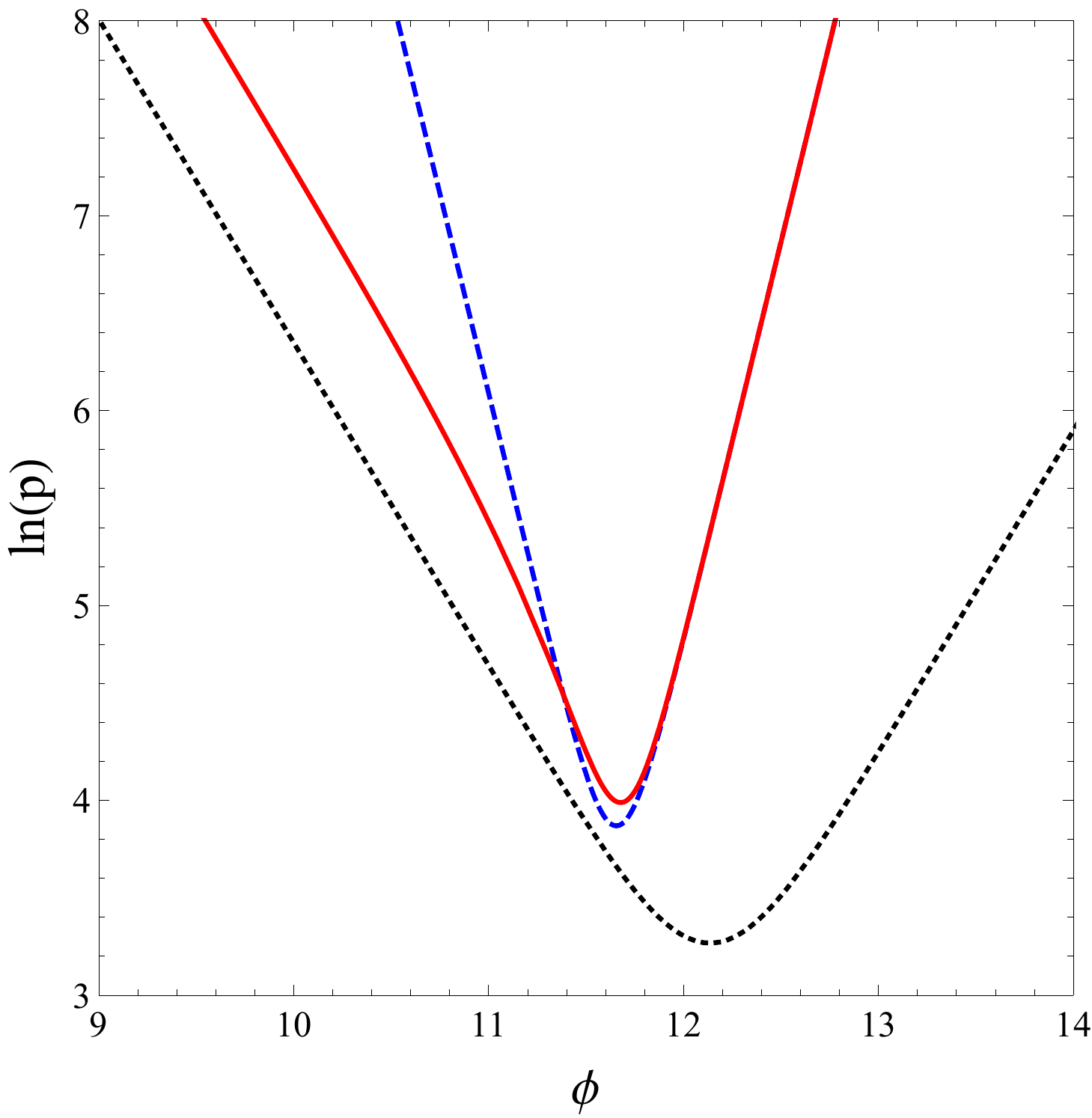}
		\end{center}
\caption{Evolution of standard triad with respect to scalar field is shown. Left plot corresponds to the $\mu_0$-scheme, and the right plot to the $\bar \mu$-scheme. The asymmetric bounce is visible in both the cases. Conventions and initial conditions for different curves are of Figs. \ref{mu0plots} and \ref{fig5} for $\mu_0$ and $\bar \mu$ cases respectively. Note the disagreement between the black-dashed curve with the red-sold curve in the pre-bounce regime. The mismatch is because of difference between the evolution of gauge-covariant triad (captured asymptotically by black-dashed curve) and the standard triad.   \label{fig3}}   
\end{figure}

In summary, it transpires that the quantum evolution through the bounce will be {\it asymmetric} by connecting two phase space points of classical/rescaled FLRW. This is confirmed by the numerical solutions for which we 
consider Barbero-Immirzi parameter $\gamma=0.2375$, and choose as initial state at late times a universe with $p(t_0)=6\times 10^4$, $\phi(t_0) = 13.5$  and $\pi_\phi(t_0)=300$ with  $t_0 =0$. From the Hamilton's equations we see that the latter value turns out to be a constant of motion. Further, the initial condition for $\phi$ plays little role because of the rescaling freedom in $\phi$ since the Hamiltonian constraint is $\phi$ independent. The corresponding initial value of $c(t_0)$ can be determined by implementing the Hamiltonian constraint. Figs. \ref{mu0plots} and \ref{fig2} show the evolution of volume, Hubble rate, connection and energy density confirming the resolution of big bang singularity which is replaced with a bounce which occurs when Hubble rate vanishes and energy density takes a maximum value. 
 The maximal energy density at the bounce is smaller compared to mainstream LQC, however in the far past/future it approaches zero or the classical spacetime. 
A pronounced difference is an asymmetric bounce when gauge covariant fluxes are included. We emphasize that this asymmetry is not caused by usage of gauge-covariant volume while plotting above figures but exists even while plotting standard triads or volumes using dynamics resulting from  (\ref{Reg_Constraint_E}). This can be seen in Fig. \ref{fig3}, where we have plotted standard triad for the same initial conditions. 

Finally, we point out that the energy density at the bounce can be made arbitrarily small by choice of scalar field momentum. As the value of $\pi_\phi$ is increased, the energy density at the bounce decreases.  As mentioned earlier, this is one of the limitations of $\mu_0$ scheme in LQC which also holds in the presence of gauge-covariant flux modifications. This limitation is overcome in the $\bar \mu$-scheme discussed in the following where energy density at the bounce turns out to be a universal maximum.

\subsection{The $\bar{\mu}$-scheme}
To overcome limitations of $\mu_0$-scheme,  the $\bar{\mu}$-scheme (also known as {\it improved dynamics}) was introduced in \cite{APS06c}. Instead of adapting the regularization parameter to the fixed minimum area-eigenvalue, the $\bar{\mu}$-scheme adapts it with the physical area $p\bar{\mu}$ of a loop with $\Delta$. The regularization choice is:
\begin{align}
\bar{\mu}:=\sqrt{\Delta/|p|}
\end{align}
While it was shown that this scheme has favorable physical properties, its implementation using LQG has not yet been established.\footnote{In the full theory the regularization parameter is chosen to be a real number $\epsilon\in\mathbb{R}$ and not dependent on the phase space variables. This allowed the replacement of physical quantities inside Poisson brackets with their regulated analogs.} As mentioned earlier, out of various possible regularizations it is only the $\bar \mu$-scheme which is known to be physically viable at ultra-violet and infra-red scales, and with physical predictions which are free from fiducial structures \cite{cs08}. 
In other words, we are forced to implement the $\bar{\mu}$-scheme {\it a posteriori} after the regularized constraint $C^\epsilon[\tilde N]$ has been obtained. This is relevant for the case of study here, as the implementation of $\epsilon\to \bar{\mu}$ and the gauge-covariant flux corrections $p\to p\;\sinc^2(c\epsilon/2)$ do not commute. In this section we will hence study the constraint which arises, if one incorporates first the gauge-covariant flux corrections, i.e. $C^{\bar{\mu}}[\tilde N]$ from (\ref{Reg_Constraint_E}), or explicitly:
\be\label{Const:MuBar}
C^{\bar{\mu}}[\tilde N]=-\frac{6}{\kappa\gamma^2\Delta}\sin^2(c\bar{\mu})\sqrt{p}^3\sinc(c\bar{\mu}/2)+\frac{\pi_\phi^2}{2\sqrt{p}^3}\sinc^{-3}(c\bar{\mu}/2)~.
\ee
\\
An important advantage of the $\bar{\mu}$-scheme in contrast to the earlier $\mu_0$ in standard LQC was that the bounce occurs at a universal value, $\rho_{\mathrm{b}} \approx 0.41 \rho_{\mathrm{Pl}}$. 
Let us see whether this feature holds when modifications due to gauge-covariant fluxes are included. Using gauge-covariant expression for energy density $\rho:=\pi_\phi^2/(2p^3 {\rm sinc}^6(\bar{\mu}c/2))$, the vanishing of the constraint (\ref{Reg_Constraint_E}) yields
\begin{align}
\rho=\frac{6}{\kappa \gamma^2\Delta}\sin^2(c\bar{\mu})\sinc^{-2}(c\bar{\mu}/2)
\end{align}
whose maximum in $0<c\bar{\mu}<\pi$ is unique and can numerically be determined at $c\bar{\mu}\simeq 1.32419$ with $\rho_{\rm max}\simeq 0.515$, which is a bigger value compared to standard LQC. Thus, unlike $\mu_0$-scheme discussed earlier, quantum bounce occurs at the same value of energy density irrespective of the initial conditions. \\

The Hamilton's equations can be derived in a straightforward way, which turn out to be (choosing lapse $\tilde N=1$),
\begin{eqnarray}
\dot{c} &=& \frac{1}{2 \gamma \Delta} \bigg[\sqrt{p} \sin^2(c \bar{\mu }) \left(\cos \left(\frac{c \bar{\mu }}{2}\right)  - 3 \, \text{sinc}\left(\frac{c \bar{\mu }}{2}\right) \right) + c \sin (2 c \bar{\mu }) \text{sinc}(\frac{c \bar{\mu }}{2}) \bigg] \nonumber \\
&& ~~~~~~~~+ \frac{\gamma \kappa  \pi_\phi^2}{8 p^{5/2} \text{sinc}^4\left(\frac{c \bar{\mu }}{2}\right)} \bigg[\cos\left(\frac{c \bar{\mu }}{2}\right) - \text{sinc}\left(\frac{c \bar{\mu }}{2}\right) - \frac{2 p^{1/2}}{\sqrt{\Delta} c}\sin \left(\frac{c \bar{\mu }}{2}\right) \bigg] \nonumber \\
&& ~~~~~~~~-\frac{p \sin \left(\frac{c \bar{\mu }}{2}\right) \sin^2\left(c \bar{\mu }\right)}{\gamma  c \Delta ^{3/2}}, \\
\dot{p}&=&\frac{1}{{16\bar{\mu} \gamma  p^{3/2} \sinc^2\left(\frac{c\bar{\mu}}{2}\right)}} \bigg[8 p^2 \cos \left(\frac{c \bar{\mu}}{2}\right) \text{sinc}^4\left(\frac{c \bar{\mu}}{2}\right) \left(c\bar{\mu}-2 \sin \left(c \bar{\mu}\right)+5 c \bar{\mu} \cos \left(c \bar{\mu}\right)\right) \nonumber \\ && ~~~~~~~~~ +\gamma ^2 \kappa \bar{\mu}^2 \pi_\phi^2 \left(c \bar{\mu}\cot \left(\frac{c \bar{\mu}}{2}\right)-2\right) \csc \left(\frac{c \bar{\mu}}{2}\right) \bigg], \\
\dot{\phi}&=&\frac{\pi_\phi}{p^{3/2}}\sinc(c\bar{\mu})^{-3}, \\
\dot{\pi}_\phi&=&0 ~.
\end{eqnarray}

The asymptotic properties can be studied in a similar way as in $\mu_0$-scheme. We are interested in two asymptotic regimes, the post-bounce regime identified by $b:= c\bar{\mu}\simeq 0$, and the pre-bounce regime corresponding to $b \simeq \pi$. In both the regimes, energy density is extremely small compared to Planck scale and one expects classical GR to hold true. 
In the regime $b\simeq 0$ we obtain the classical Friedmann and Raychaudhuri equations 
\begin{align}
\left(\frac{\dot{a}}{a}\right)^2|_{b\simeq 0}&=\tilde N^2\frac{\kappa}{6}\frac{\pi^2_\phi}{2p^3}+\mathcal{O}(p^{-4}),\\
\left(\frac{\ddot{a}}{a}\right)|_{b\simeq 0}&=-\tilde N^2\frac{\kappa}{3}\frac{\pi^2_\phi}{2p^3}+\mathcal{O}(p^{-4}) ,
\end{align}
with no change in the classical Hamilton's equations. On the other hand  in the region $b\approx\pi$, we find again in leading order in $\rho$ the same rescaling as in (\ref{Friedmann_modified}):
\begin{align}
\left(\frac{\dot{a}}{a}\right)^2|_{b\simeq \pi}&=\tilde N^2\frac{\kappa}{6}\frac{\pi^2_\phi}{2p^3}\left[\frac{\pi}{2}\right]^2+\mathcal{O}(p^{-4}), \\
\left(\frac{\ddot{a}}{a}\right)|_{b\simeq \pi}&=-\tilde N^2\frac{\kappa}{3}\frac{\pi^2_\phi}{2p^3}\left[\frac{\pi}{2}\right]^2+\mathcal{O}(p^{-4}),\\
\dot{\phi}|_{b\simeq \pi}&=\tilde N\frac{\pi_\phi}{\sqrt{p}^3}\left[\frac{\pi}{2}\right]^3+\mathcal{O}(p ^{-4}) .
\end{align}
The previous analysis can in principle be repeated for all $b:=c\bar{\mu}\simeq z\pi$ with $z\in\mathbb{Z}$. In standard LQC these choices corresponded to several branches, which were all physically indistinguishable as the flow of the Hamiltonian constraint in standard LQC interpolates between $(z-1)\pi$ and $z\pi$. In that case each point $z\pi$ corresponds to an asymptotic point where classical FLRW universe gets approached. As a result, in standard LQC it is not possible to decide the corresponding branch of the universal dynamics from physical observations.

However, in presence of the gauge-covariant fluxes the situation changes dramatically. Similar to (\ref{Friedmann_modified}), at $b\simeq z\pi$ the rescaling in the Friedmann and Raychaudhuri equations goes as $\sinc(z\pi/2)^{-2}$. In other words, the {\it only} two branches featuring as classical FLRW universe in one asymptotic limit are the two principal branches $b\in (-\pi,0)$ and $b\in (0,\pi)$. Hence, we can restrict our phase space safely to any region containing $(-\pi,\pi)$. 

\begin{figure}[tbh!]
	\begin{center}
		\includegraphics[scale=0.49]{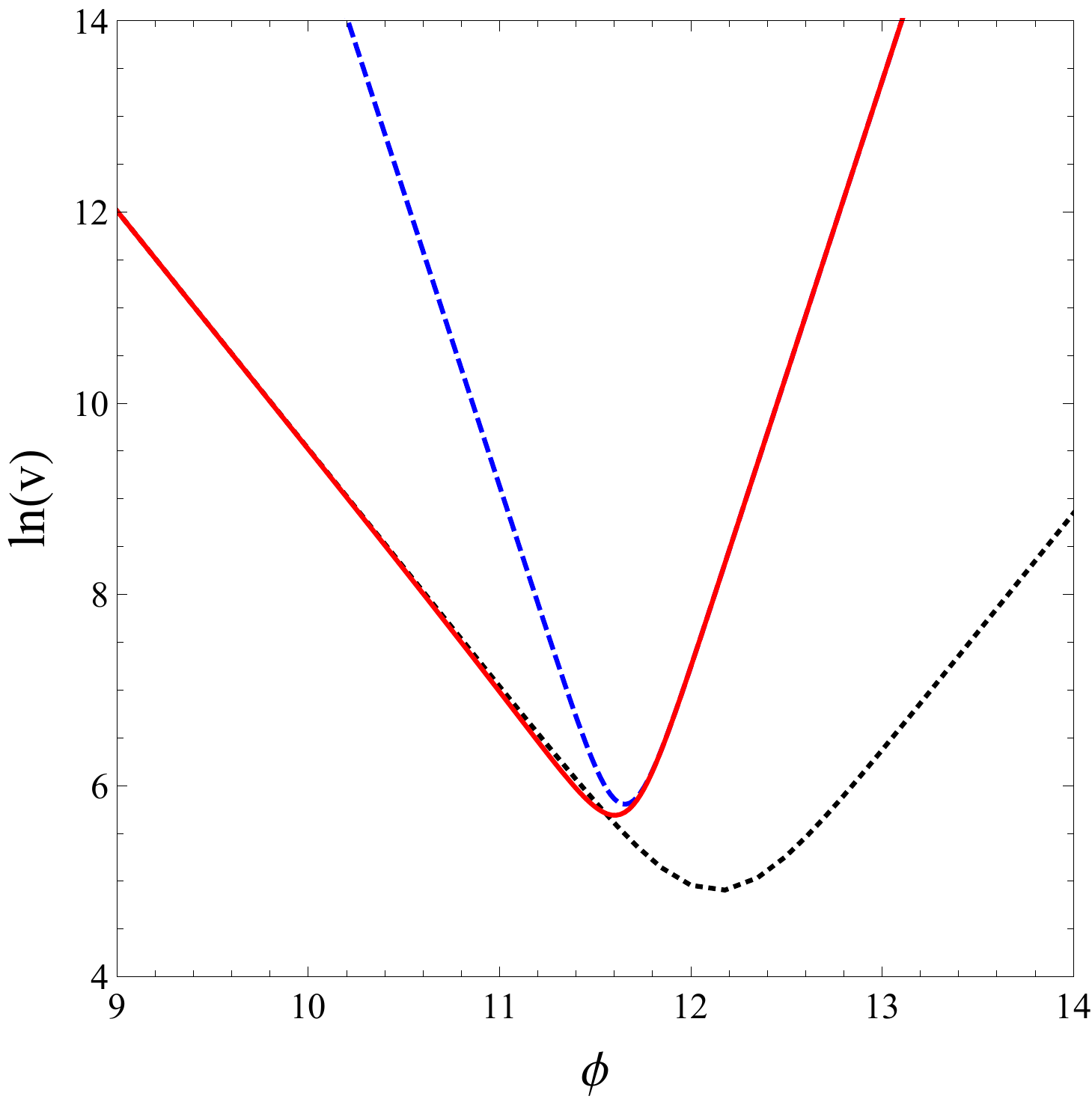}
		\includegraphics[scale=0.55]{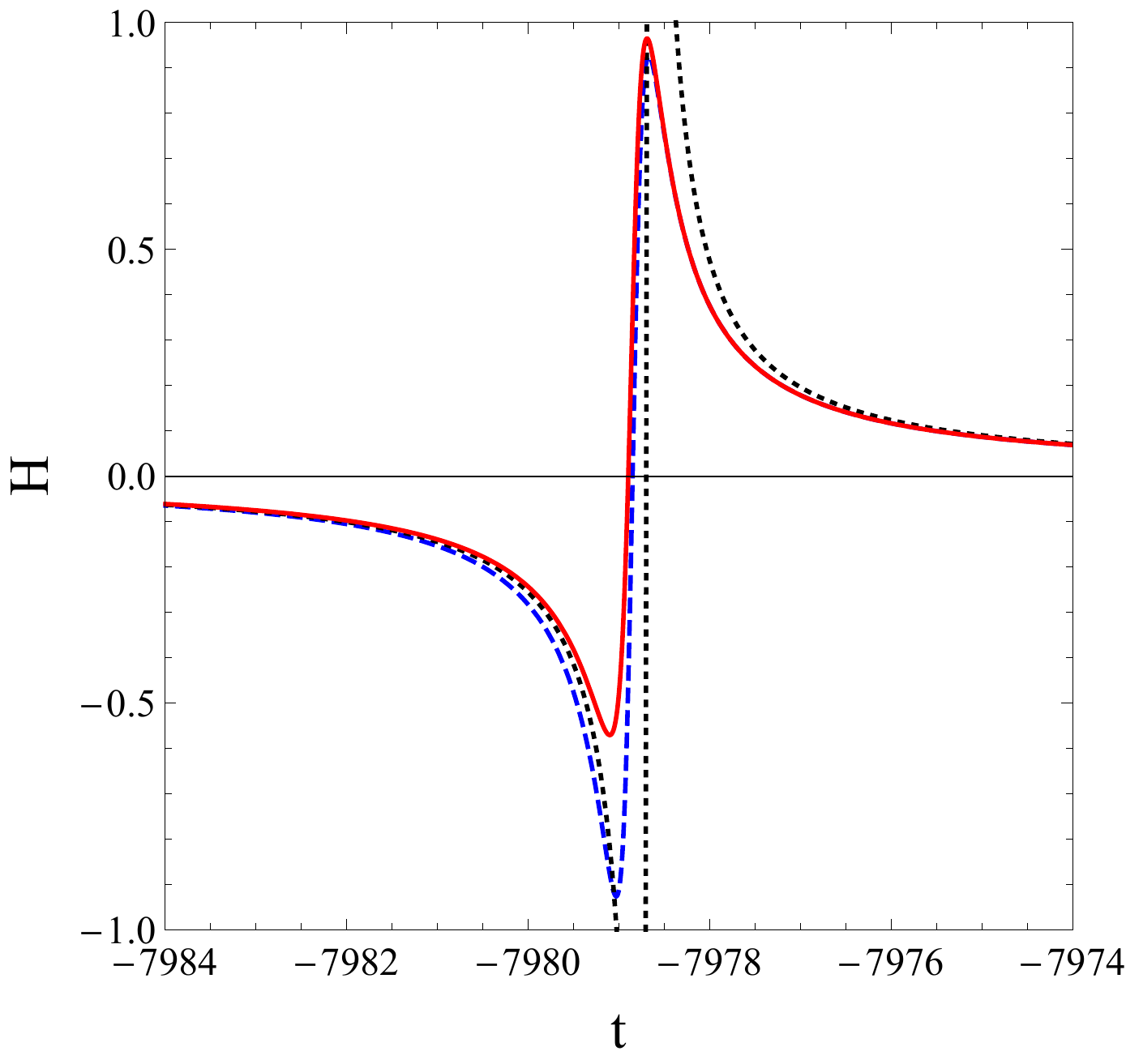}
	\end{center}
\caption{Variation of gauge-covariant volume with respect to $\phi$ and the Hubble rate with respect to $t$ are plotted for $\bar \mu$-scheme (red-solid curves). The blue-long dashed curve shows the standard LQC in the post-bounce regime, whereas the black-dashed curve shows standard LQC with rescaled values of Newton's constant and scalar field momentum (\ref{rescaling}). \label{fig5}}
\end{figure}

\begin{figure}[tbh!]
	\begin{center}
		\includegraphics[scale=0.50]{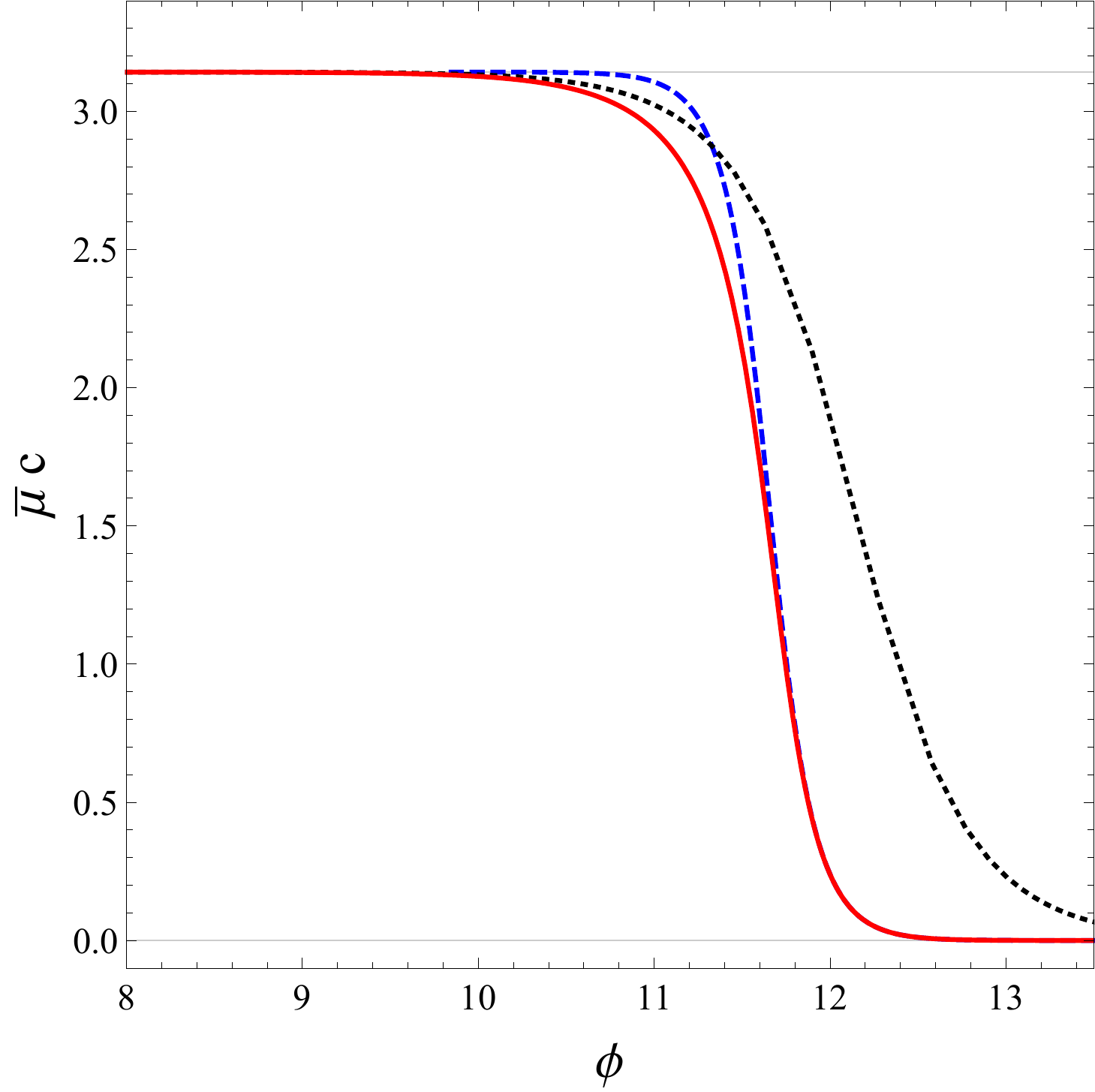}
		\includegraphics[scale=0.51]{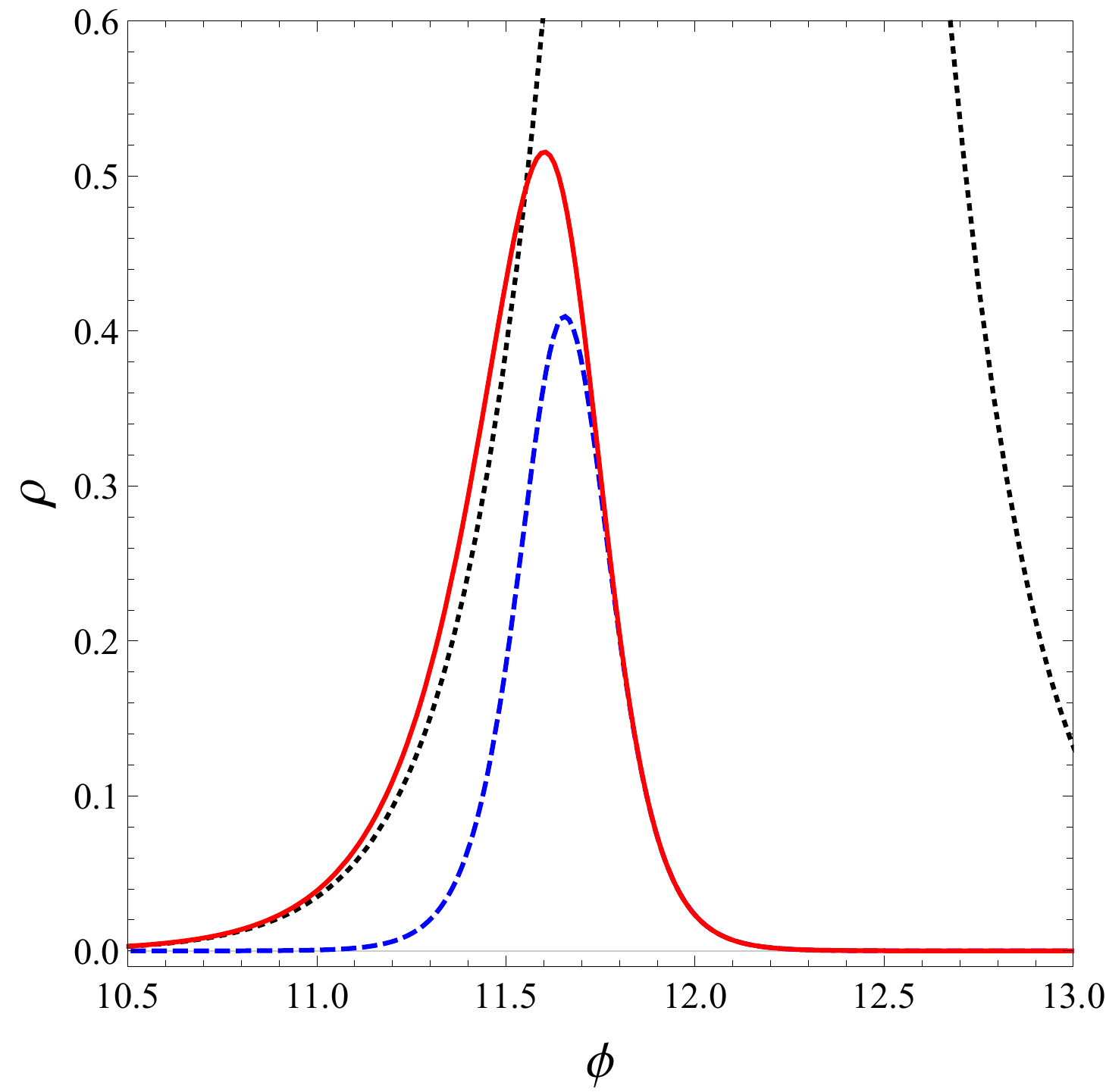}
	\end{center}
\caption{The connection $c$ and energy density $\rho$ are plotted versus $\phi$ for $\bar \mu$-scheme. Conventions are the same as in Fig. \ref{fig5}. As in standard LQC, the energy density is universally bounded but has an asymmetric profile across the bounce.  \label{fig6}}   
\end{figure}

We investigate now the flow of the scalar constraint as regularized by (\ref{Reg_Constraint_E}) numerically  as for the $\mu_0$-scheme. The following initial conditions are chosen as in the $\mu_0$-scheme: $p(t_0)=6\times 10^4$, $\phi(t_0) = 13.5$  and $\pi_\phi(t_0)=300$ where $t_0 =0$. The initial condition for $c(t_0)$ at initial time is determined via the Hamiltonian constraint.  The result is presented in Figs. \ref{fig5} and \ref{fig6} where the evolution of the  scalar constraint $C^{\bar{\mu}}$ including the gauge-covariant fluxes is compared with the  scalar constraint used in standard LQC. In these figures we have shown the variation of gauge-covariant volume, Hubble rate, $c$ and energy density, which confirm the existence of a quantum bounce in Planck regime. As in the $\mu_0$ case, the bounce is asymmetric. While LQC matches post-bounce trajectory, the pre-bounce dynamics is captured only with LQC when a rescaling of Newton's constant (\ref{rescaling}) with e.g. $\alpha=1$ is incorporated. Note that in this case the bounce always occurs at the maximum value of $\rho$. Further, as shown in Fig. \ref{fig3} the asymmetry of bounce persists even when we study standard triads for the evolution resulting from (\ref{Reg_Constraint_E}).

\begin{figure}[tbh!]
	\begin{center}
		\includegraphics[scale=0.42]{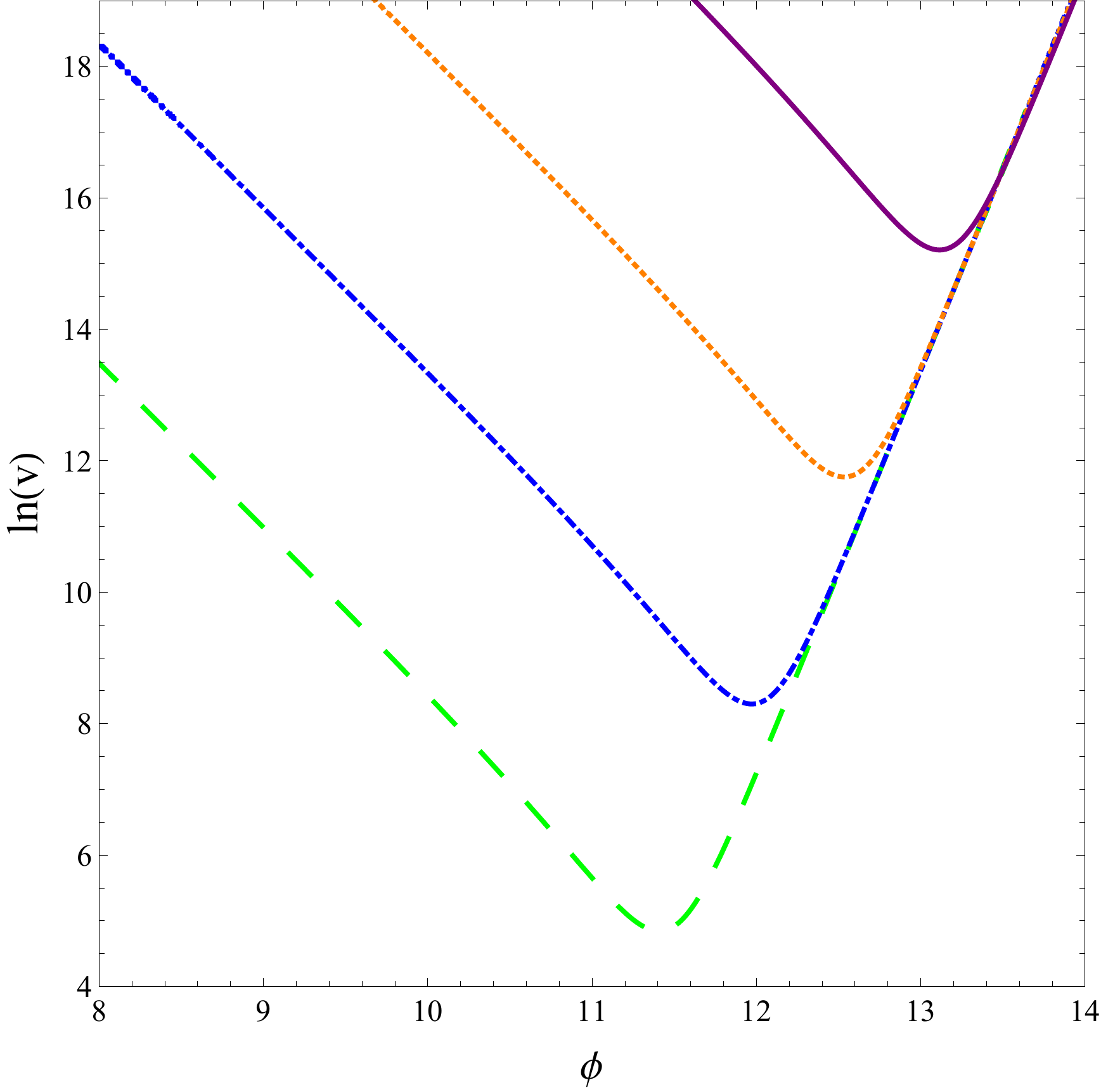}
		\includegraphics[scale=0.415]{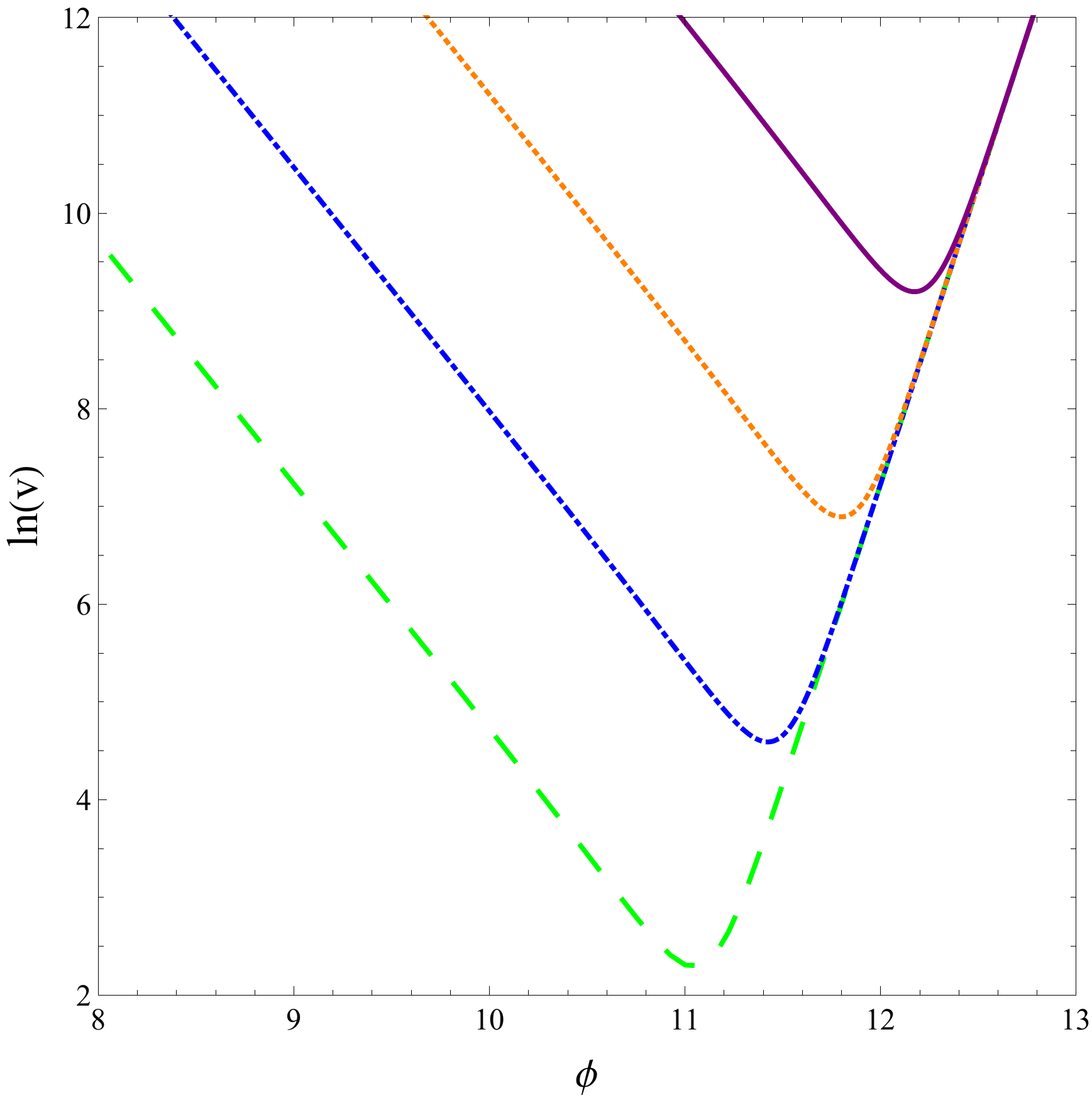}
	\end{center}
	\caption{Simulations with different initial values of $\pi_\phi$ are shown for $\mu_0$ (left) and $\bar \mu$ (right) schemes using gauge-covariant fluxes. Dashed-green curve corresponds to $\pi_\phi=10$, dot-dashed curve corresponds to $\pi_\phi=100$, $\pi_\phi=1000$ is shown in dotted-orange and $\pi_\phi=10000$ in solid purple. For all cases, we find an asymmetric bounce.\label{fig8}}   
\end{figure}

We tested robustness of our numerical results for various initial conditions and more than 500 test cases for $\mu_0$ and $\bar \mu$-schemes. The values of scalar field momentum ranged from $\pi_\phi = 10$ -- $10000$ in our simulations. Some examples of these simulations are shown in Fig. \ref{fig8}, all of which reveal asymmetric bounces. We find that the volume at the bounce depends on the value of $\pi_\phi$, with higher values of $\pi_\phi$ result in bounces at larger values of volume. 


\section{Towards LQC quantization for gauge-covariant fluxes}
\label{s4}

In this section we consider the quantization of  the Hamiltonian constraint $C^{\bar{\mu}}[N]$ from (\ref{Reg_Constraint_E}) which incorporates the  gauge-covariant flux corrections. Throughout this section the orientation of the triad will again be left unspecified. We work towards a quantization in the context of standard LQC using $\bar \mu$-scheme \cite{APS06c,acs}. A quantization using $\mu_0$ regulator can be performed following analogous steps discussed below.

Note that the gauge-covariant flux corrections introduce the appearance of terms of the form $\sin(b/2)/(b/2)$ where $b:= c \bar \mu$ and while the quantization of $\sin(b/2)$ is standard in the LQC Hilbert space in terms of shift operators, it is a priori unclear how to deal with $b^{-1}=(c\bar{\mu})^{-1}$. Hence, we will rewrite the classical expression in such a manner that it becomes suitable for quantization. 
			
It is a widely known fact that the quantization of $b$ alone is not possible, because it is not an almost periodic function. It would hence not be supported on the kinematical Hilbert space $\Hil_{kin}$ of LQC. This Hilbert space are exactly the functions with $\psi(b)=\psi(\pi-b)$ in the representation where $\sin(b)$ acts by multiplication. However, multiplication by $b$ itself, does not leave this space invariant. This fact has been encountered before in the literature, most notably in the context of Bianchi-II cosmologies. A quantization of this spacetime has been achieved in \cite{AWE09} by writing a connection operator using open holonomies which leads to  replacing  $b\mapsto \sin(b)$ before quantization. This strategy had been often used for the quantization of other cosmological space times \cite{WilEw10,CK11,CK13,SS17a,SS17b} and compared against the standard LQC quantization of the anisotropic model in \cite{SWE13}. Albeit above features of this quantization, it is not applicable in our case, as we have to incorporate an inverse power of $b$ into the framework. And as $b$ approaches zero in the far future, this would imply the quantization of a quantity which classically diverges in the present epoch.
			
Instead of considering alternative quantization schemes, one can take advantage of the fact 
that the full gauge-covariant flux correction always remains bounded: $\sinc(b/2)\leq 1$ for all $b$. It would hence be much more natural to quantize the $\sinc$-function as a whole. However, $\sinc$ is again not an almost-periodic function for all its values. Yet, classically the trajectories which belong to physical viable solutions, i.e. featuring todays universe in the far future, do not exhaust the full range of $b\in\mathbb{R}$. Indeed, as one can see from Figs. \ref{fig2} and \ref{fig6}, the argument of $\sinc(c \epsilon)$ always remains in the range $c \epsilon \in(0,\pi)$ independently of the scheme used. As outlined in the end of the last section, this together with the solution for $ c \epsilon \in (-\pi,0)$ corresponds to the principal branches. Moreover, these principal branches are the only physical viable ones, as they allow the recovery of a classical, non-rescaled FLRW universe in their asymptotic limits $b=0$. Consequently, it justifies to restrict the parameter space of $b$ to $(-\pi,\pi)$.
			
This motivates the replacement of $\sinc(b)$ by a function $TF(b)$ such that both of them agree in $(-\pi,\pi)$, while $TF(b)$ is an almost periodic function of $b$.\footnote{This is analogous to the situation in standard LQC where $b$ is considered in the range $(-\pi/\lambda, \pi/\lambda)$ (where $\lambda = \sqrt{\Delta}$) in quantization. As in LQC, this does not in any way result in ignoring any physically viable sector of the theory corresponding to evolution from late times of our universe to big bounce and beyond. This is because such a dynamical evolution corresponds to above range.} 
This can be achieved via finding the Fourier series of $\sinc$ function restricted to a compact interval $I$. We will choose $I=(-2\pi,2\pi)$ for better convergence between $\pm \pi$ at finite orders of the Fourier series. Since, $\sinc$ as well as its derivative are both continuous and square integrable, its Fourier series converges absolutely and uniformly to $\sinc$. Hence, for $b\in I$ the following holds,
\be\label{FourierSeries}
\sinc(b)^2=TF_\infty(b),\hspace{10pt}TF_N(b):=a_0/2+\sum_{n=1}^N a_n \cos(n b/2),\hspace{10pt}a_n:=\frac{1}{2\pi}\int_{-2\pi}^{2\pi}\rd x\; \frac{\sin(x)^2\cos(n x)}{x^2} ~.
\ee
Here we have taken use of the fact that $\sinc$ is an even function and hence all contributions of $\sin(nb)$ vanish.
			
\begin{figure}[tbh!]
				\begin{center}
					\includegraphics[scale=1]{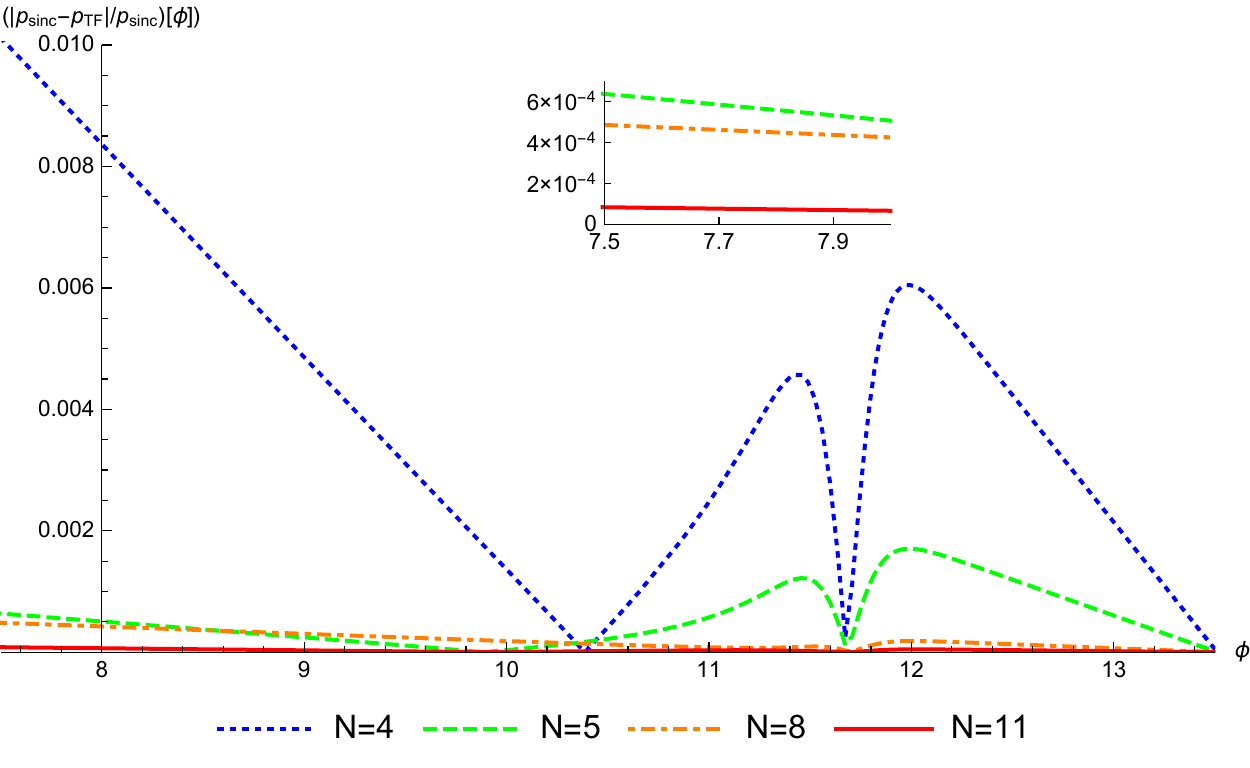}
				\end{center}
\caption{The relative error of the dynamics approximated with $TF_N$ is plotted. Starting with the same initial conditions in far future ($\phi=13.5$), we determine the evolution of a $\bar{\mu}$-Hamiltonian with $\sinc(b)$ respectively $TF_N(b)$ ($N=4,5,8,11$) as gauge-covariant-flux corrections. The canonical variable $p_{\sinc}(t)$ is well-approximated by $p_{TF}(t)$ until very late into the past. The bounce happens at $\phi= 11.5$ for $\sinc(b)$ and shortly afterwards in $TF(b)$. Outside of the quantum regime, the relative error $|p_{\sinc}-p_{TF}|/p_{\sinc}$ accumulates for $\phi\to\pm\infty$.	However, inside the region where quantum effects are important the mismatch can be made arbitrary small by choosing $N$ sufficiently large. The inset figure shows the zoom of the behavior of the curves in the far past.}
				\label{RelativeError}
\end{figure}

For practical purposes one might want to truncate the series at a finite number of terms $N$. Especially for numerical evolution of the quantum constraint a fast convergence of the series would be desirable. 
In particular, the  evolution governed by (\ref{Reg_Constraint_E}) in the $\bar{\mu}$-scheme can for example be represented by $TF_N$ with $N=4$ up to a relative error of $(|p_{\sinc}-p_{TF_N}|/p_{\sinc})[\phi]<0.01$ (see figure \ref{RelativeError}). For general estimations on the remainder term of Fourier series see for e.g. \cite{Kol34,BFLBLB07}.

Rewriting $C^{\bar{\mu}}[N]$ using $TF_N$ from (\ref{FourierSeries}) allows now to proceed to the quantization on the kinematical Hilbert space of LQC \cite{Boj99a,Boj99b,Boj05,AP11,Vel07}. 
We promote the volume $v:=p^{3/2}$ to a multiplication operator on $\Hil_{kin}:=L_2(\bar{\mathbb{R}},d\mu_{\rm Bohr}(v))$ of square integrable functions on the Bohr compactification of the real line:
\begin{align}
			\hat{V}|v\rangle=(2\pi\gamma\ell_P^2) |v|\:|v\rangle,\hspace{30pt} \langle v,v'\rangle=\delta_{v,v'}
\end{align}
where $|v\rangle$ form hence an orthonormal basis of eigenstates. Since, $b$ is the canonical conjugated variable to $v$, its exponentiation acts as a shift-operator:
\begin{align}
			\widehat{e^{i\lambda b}}|v\rangle =\hat{\mathcal{N}}^{2\lambda}|v\rangle =|v+2\lambda\rangle .
\end{align}
Note that $L_2(\bar{\mathbb{R}},d\mu_{\rm Bohr}(v))$ includes square integrable functions with negative $v$. Thus, we define as kinematical Hilbert space the subspace of symmetric states,
\begin{align}
			\Hil_{\rm gr}:=\{\psi(v)\in\Hil_{kin}\: :\: \psi(v)=\psi(-v) \}
\end{align}
by which we encode the fact that $v\to -v$ is a large gauge transformation which does not change the physics of the model \cite{APS06b}. 
				
The quantum version of $TF_N$ then simply becomes
\begin{align}
				\widehat{TF}_N:=\frac{a_0}{2} \mathds{1}_{\Hil_{\rm gr}} +\sum_{n=1}^{N}\frac{a_n}{2}  \left(\hat{\mathcal{N}}^{n}+\hat{\mathcal{N}}^{-n}\right) ~.
\end{align}
We can extend this definition to account also for $\widehat{TF}_\infty$. Indeed, this operator is well defined on $\Hil_{\rm gr}$ as it commutes with the parity operator $\Pi \psi(v):=\psi(-v)$. Moreover, it is a bounded operator of unit norm, since for $\psi(v)\in \Hil_{\rm gr}$ with $||\psi||=1$ the following holds
\begin{align}
				||\widehat{TF}_\infty\psi||=\frac{a_0}{2}||\psi||+\sum_{n=1}^{\infty}\frac{a_n}{2}(||\psi(.+n)||+||\psi(.-n)||)=\frac{1}{2}\sum_{n=-\infty}^{\infty}a_n =\sinc(0)^2=1 ~.
\end{align}
With this operator at hand, we can now finally proceed towards the quantization of $C^{\bar{\mu}}[N]$.\\
				
To keep the quantization procedure of scalar constraint (\ref{Reg_Constraint_E}) as close as possible to standard LQC, we will also incorporate the inverse volume corrections and symmetric ordering in our construction. However, we will choose as lapse function $N=v \times TF_\infty^{3/2}$ (as opposed to the usual choice of $N=v$ in solvable LQC \cite{acs}) in order to make the matter part of the constraint independent of $v$ as well as $b$. Then, we can promote the matter part to a constraint operator $\mathds{1}_{\Hil_{\rm gr}} \otimes {\rm \hat{H}}_\phi$ on the direct product Hilbert space $\Hil_{\rm total}:=\Hil_{\rm gr}\otimes \Hil_\phi$. The latter one is defined using the standard Schr\"odinger representation $\Hil_\phi=L_2(\mathbb{R},d\phi)$, on which $\hat{\phi}=\phi$ and $\hat{\pi}_\phi=i\hbar \partial_\phi$.
				
We can now immediately use known results of the literature \cite{APS06c,ADLP19}, to which we multiply the volume from the lapse and the gauge-covariant flux corrections. Note that when passing from the classical expression to operators the choice of ordering becomes important. Similar to \cite{APS06a,APS06b,APS06c} we will not only choose a symmetric ordering for the standard geometric parts of the constraint, but moreover order the sinc-terms in a symmetric fashion.
			
That is, we obtain as the full evolution operator:
\begin{align}
				-\hbar^2\partial_\phi^2=-2\;\widehat{TF}_\infty\; \sqrt{\hat{V}}\hat{C}^{\bar{\mu}}\sqrt{\hat{V}}\;\widehat{TF}_\infty\; =:\hbar^2 \Theta_{TF}
\end{align}
where the quantum scalar constraint operator is the standard LQC operator \cite{APS06c,AP11}
\begin{align}
				\hat{C}^{\bar{\mu}}|v\rangle =\frac{-3 \hbar }{32\gamma \sqrt{\Delta}}\left(f(v+2)\hat{\mathcal{N}}^4-f_0(v)\mathds{1}_{\Hil_{\rm gr}}+f(v-2)\hat{\mathcal{N}}^{-4}\right)|v\rangle
\end{align}
with
\begin{align}
				f_0(v):=f(v+2)+f(v-2),\hspace{40pt}f(v):=-|v|(|v+1|-|v-1|)
\end{align}
It is now of interest, to study further properties of the finite difference operator $\Theta_{TF}$ and the quantum evolution it produces and its comparison with the regularized dynamics. We will come back to this task in a later publication.\\

\section{Conclusion}
\label{s5}

In the last decade and half progress in LQC has provided a promising  avenue to understand quantum gravitational effects for cosmological models. Resolution of cosmological singularities for various spacetimes has been explored, along with departures from GR in the very early universe with potential signatures in CMB. However, until today the connection between LQC and LQG is still unknown. As a result, the way Planck scale physics in LQC changes as we bring it closer to methods of LQG is an important question to be answered to understand robustness and reliability of phenomenological predictions. One way to answer these questions would be to explore the cosmological sector of LQG which has recently attracted a lot of activity \cite{BEHM16,BEHM17,EV18,AC12,AC14a,ABLS19,DL17a,DL17b}. A preliminary insight from these investigations is that the physics, at least of singularity resolution, should change from standard LQC \cite{AC12,AC14a,ABLS19,DL17a,DL17b}. Here coherent state methods have proved to be very useful to explore cosmological implications of LQG and compare with LQC. Despite these advances there are serious gaps in these constructions, since some aspects of the semiclassical analysis of coherent states on fixed graphs in LQG are so far not projected to LQC. An example is that the commutator between right-invariant vector fields is non-zero, whereas the classical Poisson-bracket between triads vanishes. One way to solve this puzzle comes from using gauge-covariant fluxes \cite{LS19a}, introduced first by Thiemann \cite{ThiVII00}, which leads to 
classical non-commuting phase space functions and is well suited for coherent state techniques in LQG \cite{Winkler1,Winkler2,Winkler3,ThiemanComplex,SahThiWin}.

In this manuscript we have introduced a new regularization scheme for the constraint operator of LQC which is based on switching from the standard fluxes, i.e. the electric field smeared against two-dimensional surfaces, to said gauge-covariant fluxes following Thiemann's construction \cite{ThiVII00}. In contrast to former proposals, the new fluxes transform feasibly under gauge transformations and allow a straightforward construction of gauge-invariant quantities in the presence of finite discretization. One such possible discretization is a cubic lattice with finite lattice spacing. For this, we repeated the construction of the gauge-covariant fluxes for cosmology and have shown an actual regularization to the volume of a region, such that it is (i) build only from the discrete lattice phase space variables, (ii) reduces to the actual volume of the region in the limit of vanishing regularization parameter, and (iii) gauge-invariant for all values of the regularization parameter. 
An advantage of this procedure is that it simplifies computations in the full theory as it rightfully allows to skip projecting the gauge coherent states to the Gauss-invariant Hilbert space. One can take use of the fact that the expectation value of gauge-invariant quantities for group-averaged coherent states (which are solutions to the quantum Gauss constraint) equals in leading order in the spread with the coherent states, before the averaging procedure. Hence, motivating regularizations for LQC from the full theory becomes a rigorous procedure in our setting.

Then, we have studied the form of these gauge-covariant flux corrections to the Hamiltonian constraint of spatially-flat, isotropic cosmological spacetime sourced with a massless scalar field. In the scalar constraint we considered, the Lorentzian term is combined with the Euclidean term owing to classical symmetry reduction as in standard LQC before any discretization is introduced. Modifications where Euclidean and Lorentzian terms are treated independently, along with inclusion of cosmological constant have also been studied in a companion paper \cite{LS19c}. We find that the Hamiltonian constraint is modified with additional bounded sinc-term of the connection emerging from the transformation $p\to p\;\sinc^2(c\epsilon/2)$. This modification affects both the gravitational and matter parts of the scalar constraint. Since the modifications depend on connection, and hence spacetime curvature, a novel change from standard LQC is that matter behaves as non-minimally coupled. This can have interesting phenomenological consequences including for inflationary spacetimes which will be explored elsewhere.

 In the present manuscript, we have focused our attention on the non-trivial changes in the physics of the quantum bounce originating from the use of gauge-covariant fluxes for $\mu_0$ as well as $\bar \mu$ schemes. We show that 
in both schemes there exists a transition through the quantum region in form of a bounce between our universe in the far future and an old universe in the far past. Once, again this presents a resolution of the initial singularity. However, in contrast to mainstream LQC the evolution as driven by the standard regularization of the scalar constraint is no longer symmetric, i.e. the universe in the far past can be matched to a classical contracting one with modified gravitational coupling constant.\footnote{Indeed, there exists a 1-parameter family of classical contracting FLRW solutions with different rescalings moreover for scalar field
momentum and lapse function. However, both of these values are pure gauge and their rescaling is therefore not observable.} Asymptotic properties of pre-bounce and post-bounce solutions show that the bounce is generically asymmetric and there exists no values of $G$ such that there is no rescaling of effective constants across the bounce. This situation holds true for $\mu_0$ as well as $\bar \mu$ scheme. The asymmetric bounce is thus an inherent feature of Hamiltonian with gauge-covariant flux modifications. Our analysis provides a concrete example of rescaling of constants across the singularity, a phenomena speculated earlier in \cite{smolin} and studied for discrete quantum gravity \cite{pullin} as well as for Thiemann regularization of LQC using triads \cite{LSW18a,ADLP19}. A difference from earlier studies of \cite{smolin,pullin} is that the change of constants is not random but completely fixed by the dynamics. 

Let us note that  regularized dynamics studied in this manuscript is at the moment not proven to be the effective dynamics of a corresponding quantum cosmology theory. In a sense, our treatment is similar to various works in standard LQC where effective dynamics is often assumed to understand quantum gravitational implications. To address this question, we proposed a procedure by which the gauge-covariant flux corrections can be promoted to an operator on the physical Hilbert space of LQC. The strategy outlined in this paper, will also prove vital for the quantization of other regularizations for the scalar constraint, such as when the Lorentzian part is treated independently in presence of gauge-covariant fluxes \cite{LS19c}. Based on the observation that sinc is a bounded function, it was on the classical phase space replaced with its corresponding Fourier-series on a compact region. This region was chosen big enough that it incorporates all possible phase space trajectories, which at some point correspond to cosmological dynamics potentially relevant for our universe. This leads to a quantum evolution operator $\Theta_{TF}$ which consists of an infinite sum over shifts. In this sense the evolution operator is non-local on the LQC lattice and differs from the one in standard LQC because of higher order quantum difference operators. Though the quantum evolution operator we proposed is technically more involved than in standard LQC, we should note that 
the contributions of high lattice distances are exponentially fast dropping of with the distance on the lattice. Similar situations are already known to the literature, e.g. in the context of perfect actions for quantum field theories \cite{Has98,Has08,LLT3}. 

Our analysis opens a new window to incorporate further techniques from LQG to cosmological spacetimes by incorporating gauge-covariant fluxes and in this sense providing a first ever SU(2) gauge-invariant treatment of singularity resolution using LQG techniques. It results in a striking change from the existing results in LQC. The symmetric bounce in simplest models is replaced by an asymmetric bounce with a change in effective constants in the pre-bounce regime. It remains to be seen how this change affects the physics of the very early universe and the potential signatures in CMB. Moreover, the theoretical techniques used here can be further refined and generalized. Examples of these include using graph-coherent states \cite{Mehdi} or stable coherent states \cite{Antonia}.
  Finally, one can hope that these insights into the cosmological model help to deal with the vast regularization ambiguities of the full theory. Some of these ambiguities and their physical implications  are studied in our companion paper \cite{LS19c}. 

\section*{Acknowledgements}
We thank Thomas Thiemann for several discussions. This work is supported by NSF grant PHY-1454832.


\begin{thebibliography}{99}
{\footnotesize	\setlength{\parskip}{0.0em}
\setlength{\itemsep}{0.6pt}
	
\bibitem{ThiVII00}
	T. Thiemann,
	Quantum Spin Dynamics (QSD): VII. Symplectic Structures and Continuum Lattice Formulations of Gauge Field Theories,
	{\it Class. Quant. Grav.} {\bf 18}, 3293-3338
	[arXiv:hep-th/00052232]
	(2001)
	
	
\bibitem{Rov04}
	C. Rovelli,
	Quantum Gravity,
	{\it Cambridge University Press}
	(2004)
	
\bibitem{AL04}
	A. Ashtekar, J. Lewandowski,
	Background independent quantum gravity: A Status report,	
	{\it Class. Quant. Grav.}
	{\bf 21}, R53-R152
	(2004)
	
\bibitem{Thi07}
	T. Thiemann,
	Modern Canonical Quantum General Relativity,
	{\it Cambridge University Press}
	(2007)
	
\bibitem{ADM62}
	R. Arnowitt, S. Deser, C. Misner,
	The Dynamics of General Relativity,
	{\it In:} Gravitation: An introduction to current research by L Witten (ed) 
	{\it New York}
	227-265
	(1962)
	
\bibitem{Ash86}
	A. Ashtekar,
	New variables for classical and Quantum Gravity,
	{\it Phys. Rev. Lett.}{\bf 57}, 2244-2247
	(1986)
	
\bibitem{Ash87}
	A. Ashtekar,
	New Hamiltonian formulation of General Relativity,
	{\it Phys. Rev. D} {\bf 36}, 1587-1602
	(1987)
	

\bibitem{Bar94}
	J.F. Barbero,
	A real polynomial formulation of General Relativity in terms of connection,
	{\it Phys. Rev. D} {\bf 49}, 6935-6938
	(1994)
	




\bibitem{Boj05}
	M. Bojowald,
	Loop Quantum Cosmology,
	{\it Living Rev. Relativity} {\bf 8}, 11
	(2005)


\bibitem{AP11}
	A. Ashtekar, P. Singh,
	Loop Quantum  Cosmology: A Status Report,
	{\it Class. Quant. Grav.} {\bf 28}, 213001
	(2011)

	
\bibitem{APS06a}
	A. Ashtekar, T. Pawlowski, P Singh,
	Quantum Nature of the Big Bang,
	{\it Phys. Rev. Let.} {\bf 96}, 141301
	(2006)

\bibitem{APS06b}
	A.	Ashtekar, T. Pawlowski, P. Singh,
	Quantum Nature of the Big Bang: An Analytical and Numerical Investigation,
	{\it Phys. Rev. D} {\bf 73}, 124038
	(2006)
	
\bibitem{acs}    A.~Ashtekar, A.~Corichi and P.~Singh,
Robustness of key features of loop quantum cosmology,
{\it Phys.\ Rev.\ D} {\bf 77}, 024046 (2008)

\bibitem{as-review} For a recent review of these developments see,     I. Agullo, P. Singh, {\it{Loop Quantum Cosmology,}} 
in {\it{Loop Quantum Gravity:  The First 30 Years,}}  Eds:   A.  Ashtekar, J.   Pullin,   World   Scientific   (2017)	



\bibitem{gop} R.~Gambini, J. Omedo and J.~Pullin,  Quantum black holes in loop quantum gravity, {\it Class. Quant. Grav.} \textbf{31}, 095009 (2014).

\bibitem{aos} A.~Ashtekar, J.~Olmedo and P.~Singh, Quantum extension of the Kruskal spacetime,
{\it Phys.\ Rev.\ D} {\bf 98},  126003 (2018)




\bibitem{BF07}
	J. Brunnemann. C Fleischhack,
	On the Configuration Spaces of Homogeneous Loop Quantum Cosmology and Loop Quantum Gravity,
	[arXiv:0709.1621]
	(2007)
	
\bibitem{Fle15}
	C. Fleischhack,
	Kinematical Foundations of Loop Quantum Cosmology,
	[arXiv:1505.04400]
	(2015)
	
\bibitem{BEHM16}
	C. Beetle, J. Engle, M. Hogan, P. Mendonca,
	Diffeomorphism invariant cosmological symmetry in full quantum gravity,
	{\it Int. J. Mod. Physics D} {\bf 25}, 1642012
	(2016)
	
\bibitem{BEHM17}
	C. Beetle, J. Engle, M. Hogan, P. Mendonca,
	Diffeomorphism invariant cosmological sector in loop quantum gravity,
	[arXiv:1706.02424]
	(2017)
	
\bibitem{EV18}
	J. Engle, I. Vilensky,
	Deriving loop quantum cosmology dynamics from diffeomorphism invariance,
	[arXiv:1802.01543]
	(2018)
	
\bibitem{AC12}
	E. Alesci, F. Cianfrani,
	A new perspective on cosmology in Loop Quantum Gravity,
	[arXiv:1210.4504]
	(2012)
	
\bibitem{AC14a}
	E. Alesci, F. Cianfrani,
	Quantum Reduced Loop Gravity: Semiclassical limit,
	[arXiv:1402.3155]
	(2014)

	
\bibitem{ABLS19}
	E. Alesci, G. Botta, G. Luzi, G Stagno,
	Bianchi I effective dynamics in Quantum Reduced Loop Gravity,
	[arXiv:1901.07140]
	(2019)
	
\bibitem{DL17a}
	A. Dapor, K. Liegener,
	Cosmological Effective Hamiltonian from full Loop Quantum Gravity,
	{\it Phys. Lett. B} {\bf 785}, 506-510
	(2018)
	
\bibitem{DL17b}
	A. Dapor, K. Liegener,
	Cosmological Coherent State Expectation Values in LQG I. Isotropic Kinematics,
	{\it Class. Quant. Grav. }{\bf 35}, 135011
	[arXiv:1710.04015]
	(2018)
	
\bibitem{LL19}
	K. Liegener, R. Lukasz,
	Cosmological Coherent State Expectation Values in LQG II. Thiemann-regularized Hamiltonian,
	{\it (to appear)}
	


\bibitem{ACZ98}	
	A. Ashtekar, A. Corichi, J. Zapata,
	Quantum theory of geometry: III. Non-commutativity of Riemannian structures,
	{\it Class. Quant. Grav.} {\bf 15}, 2955-2972
	(1998)

\bibitem{FL04}
	L. Freidel, D. Louapre,
	Ponzano-Regge model revisited I: Gauge fixing, observables and interacting spinning particles,
	{\it Class. Quant. Grav.} {\bf 21},  5685-5726
	(2004)

\bibitem{BDOT10}
	A. Baratin, B. Dittrich, D. Oriti, J. Tambornino,
	Non-commutative flux representation for loop quantum gravity,
	{\it Class. Quant. Grav. }{\bf 28}, 175011
	(2010)
	
\bibitem{FGZ11}
	L. Freidel, M. Geiller, J. Ziprick,
	Continuous formulation of the Loop Quantum Gravity phase space,
	{\it Class. Quant. Grav.} {\bf 30}, 085013 
	(2011)
	
\bibitem{BD13}
	V. Bonzom, B. Dittrich,
	Dirac's discrete hypersurface deformation algebras,
	[arXiv:1304.5983]
	(2013)
	
\bibitem{DG14}
	B. Dittrich, M. Geiller,
	A new vacuum for Loop Quantum Gravity,
	[arXiv:1401.6441]
	(2015)
	
\bibitem{CP16}
	A. S. Cattaneo, A. Perez,
	A note on the Poisson bracket of 2d smeared fluxes in loop quantum gravity,
	[arXiv:1611.08394]
	(2016)
	
\bibitem{DFG17}
	M. Dupuis, L. Freidel, F. Girelli,
	Discretization of 3d gravity in different polarizations,
	{\it Phys. Rev. D} {\bf 96}, 086017
	(2017)
	
\bibitem{FGS18}
	L. Freidel, F. Girelli, B Shoshany,
	2+1D Loop Quantum Gravity on the Edge,
	[arXiv:1811.04360]
	(2018)
	
\bibitem{MMP10}
	E. Magliaro, A. Marciano, C. Perini,
	Coherent states for FLRW space-times in loop quantum gravity,
	{\it Phys.Rev. D} {\bf 83}, 044029
	(2011)
	
\bibitem{LS19a} K. Liegener, P. Singh, Gauge-invariant bounce from quantum geometry,  arXiv:1906.02759 
	
\bibitem{Winkler1}
	T. Thiemann,
	Gauge Field Theory Coherent States (GCS): I. General Properties,
	{\it Class. Quant. Grav.}
	{\bf 18}, 2025-2064
	(2001)

\bibitem{Winkler2}
	T. Thiemann, O. Winkler,
	Gauge Field Theory Coherent States (GCS): II. Peakedness Properties,
	{\it Class. Quant. Grav.}, 2561-2636
	{\bf 18}
	(2001)

\bibitem{Winkler3}
	T. Thiemann, O. Winkler,
	Gauge Field Theory Coherent States (GCS): III. Ehrenfest Theorems,
	{\it Class. Quant. Grav.},  4629-4682
	{\bf 18}
	(2001)

\bibitem{ThiemanComplex}
	T. Thiemann,
	Complexifier  coherent  states  for  quantum  general  relativity,
	{\it Class. Quant. Grav.}
	{\bf 23}, 2063-2118
	(2006)

\bibitem{SahThiWin}
	H. Sahlmann, T. Thiemann, O. Winkler,
	Coherent states for canonical quantum General Relativity and the infinite tensor product extension,
	{\it Nucl. Phys. B}
	{\bf 606}, 401-440
	(2001)	
	
\bibitem{AQG2}
	K. Giesel, T. Thiemann,
	Algebraic Quantum Gravity (AQG) II. Semiclassical Analysis,
	{\it Class. Quant.Grav. } {\bf 24} 2499-2564
	(2007)
	

\bibitem{abl} A.~Ashtekar, M.~Bojowald and J.~Lewandowski,
  Mathematical structure of loop quantum cosmology,
  Adv.\ Theor.\ Math.\ Phys.\  {\bf 7},  233 (2003)

\bibitem{APS06c}
	A. Ashtekar, T. Pawlowski, P. Singh,
	Quantum Nature of the Big Bang: Improved dynamics,
	{\it Phys. Rev. D} {\bf 74}, 084003
	(2006)
	
\bibitem{LS19c}
K. Liegener, P. Singh,
Some physical implications of regularization ambiguities in SU(2) gauge-invariant loop quantum cosmology [arXiv:1908.07543] (2019)

	

\bibitem{RS94}
	C. Rovelli, L. Smolin,
	Discreteness of area and volume in quantum gravity,
	{\it Nucl. Phys. B} {\bf 442}, 593-622
	(1995)

\bibitem{AL96}
	A. Ashtekar, J. Lewandowski,
	Quantum Theory of Gravity I: Area Operators,
	{\it Class. Quant. Grav.} {\bf 14}, A55-A82
	(1996)
	
\bibitem{AL98}
	A. Ashtekar, J. Lewandowski,
	Quantum Theory of Geometry II: Volume Operators,
	{\it Adv. Theor. Math. Phys} {\bf 1}, 388-429
	(1998)
	
\bibitem{Thi98a}
	T. Thiemann,
	Quantum Spin Dynamics (QSD) I,
	{\it Class. Quant. Grav.} {\bf 15}, 839-873
	(1998)
	
\bibitem{Thi98b}
	T. Thiemann,
	Quantum Spin Dynamics (QSD) II,
	{\it Class. Quant. Grav.} {\bf 15}, 875-905
	(1998)

\bibitem{KS75}
J. Kogut, L Susskind,
Hamiltonian Formulation of Wilsons Lattice Gauge Theories,
{\it Phys. Rev. D} {\bf 11}
(1975)

\bibitem{cs08} 
A.~Corichi, P.~Singh,
Is loop quantization in cosmology unique?,
{\it Phys.\ Rev.\ D} {\bf 78}, 024034 (2008)



\bibitem{numlsu-1}
P. Diener, B. Gupt, P. Singh,
Numerical simulations of a loop quantum cosmos: robustness of the
quantum bounce and the validity of effective dynamics,
{\it Class. Quant. Grav.} {\bf 31}, 105015 (2014)

\bibitem{numlsu-2}
P. Diener, A. Joe, M. Megevand, P. Singh,
Numerical simulations of loop quantum Bianchi-I spacetimes,
{\it Class. Quant. Grav.} {\bf 34}, 094004 (2017)


\bibitem{Taveras}
	V. Taveras,
	Corrections to the Friedmann Equations from LQG for a
	Universe with a Free Scalar Field.
	{\it Phys. Rev. D} {\bf 78}, 064072
	(2008)
	
\bibitem{LSW18a}
B. Li, P. Singh, A. Wang,
Towards Cosmological Dynamics from Loop Quantum Gravity,
{\it Phys Rev. D} {\bf 97}, 084029 
(2018)

%
%



\bibitem{AWE09}	
	A. Ashtekar, E. Wilson-Ewing,
	Loop quantum cosmology of Bianchi type II models,
	{\it Phys. Rev. D} {\bf 80}, 123532
	(2009)
	
\bibitem{WilEw10}
	E. Wilson-Ewing,
	Loop quantum cosmology of Bianchi type IX models,
	{\it Phys. Rev. D} {\bf 82}, 043508
	(2010)
	
\bibitem{CK11}
	A. Corichi, A. Karami,
	Loop quantum cosmology of k=1 FRW: A tale of two bounces,
	{\it Phys. Rev. D} {\bf 84}, 044003
	(2011)
	
\bibitem{CK13}
	A. Corichi, A. Karami,
	Loop quantum cosmology of k=1 FLRW: Effects of inverse volume corrections,
	{\it Class. Quant. Grav.} {\bf 31}, 035008
	(2014)
	
%
%

\bibitem{SS17a} 
	S. Saini, P. Singh,
	Generic absence of strong singularities in loop quantum Bianchi-IX spacetimes,
	 {\it Class.\ Quant.\ Grav.}\  {\bf 36}, 105014 (2019)
	
	
\bibitem{SS17b}	S.~Saini and P.~Singh,
  Von Neumann stability of modified loop quantum cosmologies,
{\it  Class.\ Quant.\ Grav.}\  {\bf 36}, 105010 (2019)
	
	
\bibitem{SWE13}
	P. Singh, E Wilson-Ewing,
	Quantisation ambiguities and bounds on geometric scalars in anisotropic loop quantum cosmology,
	{\it Class. Quant. Grav.} {\bf 31}, 035010
	(2014)
	
\bibitem{Kol34}
	A Kolmogoroff,
	Zur Groessenordnung des Restgliedes Fourierscher Reihen Differenzierbarer Funktionen,
	{\it Ann. of Math.} {\bf 36}, 521-526
	(1935)
	
\bibitem{BFLBLB07}	
	C. Barrera-Figueroa, A. Lucas-Bravo, J. Lopez-Bonilla,
	The remainder term in Fourier series and its relationship with the Basel problem,
	{\it Ann. Math. et Info.} {\bf 34}, 17-28
	(2007)

\bibitem{Boj99a}
	M. Bojowald,
	Loop Quantum Cosmology I: Kinematics,
	{\it Class. Quant. Grav.} {\bf 17}, 1489-1508
	(1999)
	
\bibitem{Boj99b}
	M. Bojowald,
	Loop Quantum Cosmology II: Volume Operators,
	{\it Class. Quant. Grav.} {\bf 17}, 1509-1526
	(1999)

\bibitem{Vel07}
	J. Velhino,
	The Quantum Configuration Space of Loop Quantum Cosmology,
	{\it Class Quant Grav.} {\bf 24}, 3745-3758
	(2007)

\bibitem{ADLP19}
	M. Assanioussi, A. Dapor, K. Liegener, T. Pawlowski,
	Emergent de Sitter epoch of the quantum Cosmos: a detailed analysis.
	[arXiv:1906.05315]
	(2019)
	
	

\bibitem{smolin}  L. Smolin, Did the Universe evolve? {\it Class. Quant. Grav.} {\bf 9}, 173 (1992)

\bibitem{pullin} R.~Gambini, J.~Pullin,
Discrete quantum gravity: A Mechanism for selecting the value of fundamental constants,
{\it Int.\ J.\ Mod.\ Phys.\ D} {\bf 12}, 1775 (2003)




\bibitem{Has98}
	P. Hasenfratz,
	Prospects for perfect actions,
	{\it Nucl. Phys. Proc. Suppl.} {\bf 63}, 53-58
	(1998)
	
\bibitem{Has08}
	P. Hasenfratz,
	The theorectical background and properties of perfect actions,
	[arXiv:hep-lat/9803027]
	(2008)

\bibitem{LLT3}
	T. Lang, K. Liegener, T. Thiemann,
	Hamiltonian Renormalisation III: Renormalisation Flow of 1+1 dimensional scalar fields: Properties,
	{\it Class. Quant. Grav.} {\bf 35}, 245013
	(2018)

\bibitem{Mehdi}
	M. Assanioussi,
	Polymer quantization of connection theories: Graph coherent states,
	{\it Phys. Rev. D} {\bf 98}, 045016 
	(2018)
	
\bibitem{Antonia}
	T. Thiemann, A. Zipfel,
	Stable Coherent States,
	{\it Phys. Rev. D} {\bf 93}, 084030 
	(2016)
	
%
%
%
%
%
%
%
%
%
%
%
%

%
%
%

%
%
}	
\end{thebibliography}
\end{document}